\documentclass[a4paper,12pt]{article}

\usepackage{latexsym}

\newcommand{\ud}{\mathrm{d}}

\newcommand\omet{\tilde{g}}

\newcommand\oimet{\tilde{g}}

\newcommand\oric{\tilde{R}}

\newcommand\osric{\tilde{S}}

\newcommand\oein{\tilde{G}}

\newcommand\onab{\tilde{\nabla}}

\newcommand\Told{\tilde{T}}

\newcommand\obox{\stackrel{\sim}{\Box}\!\!}

\begin{document}

\title{Nonlinear massive spin--two field generated by higher derivative gravity} 

\author{Guido MAGNANO \\ 
Dipartimento di Matematica,
Universit\`a di Torino, \\ 
via Carlo Alberto 10, 10123 Torino, Italy \\ and \\
Leszek M. SOKO\L{}OWSKI \\ 
Astronomical Observatory, 
Jagellonian University, Orla 
171, \\ Krak\'ow 30-244, Poland}

\date{}\maketitle

\begin{abstract}
We present a systematic exposition of the Lagrangian field theory for the massive spin--two field
generated in higher--derivative gravity upon reduction to a second--order theory
by means of the appropriate Legendre transformation. It has been noticed by various authors that this nonlinear
field overcomes the well known inconsistency of the theory for a linear massive spin--two field interacting with
Einstein's gravity. Starting from a Lagrangian quadratically
depending on the Ricci tensor of the metric, we explore the two possible
second--order pictures usually called ``(Helmholtz--)Jordan frame'' and
``Einstein frame''. In spite of their mathematical equivalence, the two
frames have different structural properties: in Einstein frame, the
spin--two field is minimally coupled to gravity, while in the other frame
it is necessarily coupled to the curvature, without a separate kinetic
term. We prove that the theory admits a unique and linearly stable ground
state solution, and that the equations of motion are consistent, showing
that these results can be obtained independently in either frame (each
frame therefore provides a self--contained theory). The full equations of
motion and the (variational) energy--momentum tensor for the spin--two
field in Einstein frame are given, and a simple but nontrivial exact
solution to these equations is found. The comparison of the
energy--momentum tensors for the spin--two field in the two frames suggests
that the Einstein frame is physically more acceptable. We point out that
the energy--momentum tensor generated by the Lagrangian of the linearized
theory is unrelated to the corresponding tensor of the full
theory. It is then argued that the  ghost--like nature of the nonlinear
spin--two field, found long ago in the linear approximation, may not be so
harmful to classical stability issues, as has been expected. 
\end{abstract} 


\section{Introduction}
A consistent theory of a gravitationally interacting spin--two field
could not be developed until a significant progress was made in an
apparently unrelated subject, i.e.~higher--derivative metric theories of
gravity. It is well known that a single linear spin--two field cannot be
consistently coupled to gravity. It is therefore a common belief that
Nature avoids the consistency problem by simply not creating fundamental
spin--two (nor higher spin) fields except gravity itself. Nevertheless
the subject has remained fascinating over decades and some authors have
studied various aspects of linear spin--two fields \cite{T,NN,DW}, in
particular their dynamics in Einstein spaces \cite{Ben,DW2}.

On the other hand, higher--derivative metric theories of gravity, where
the Lagrangian is a scalar nonlinear function of the curvature tensor
(hence in this paper they are named nonlinear gravity theories, NLG) have
attracted much more attention. Most work was centered on quadratic
theories, i.e.~on Lagrangians being quadratic polynomials in the Ricci
tensor and the curvature scalar \cite{NLG,AEL}, but several authors
studied more general Lagrangians \cite{Gen}. These theories turned out to
be inadequate as candidates for foundations of quantum gravity since they
are non--unitary, but recently play a role as effective field theories.
What is more relevant here, it was found that their particle spectrum
contains a massive spin--two field. The dynamics of this field can be 
described and investigated by recasting the fourth-order NLG theory into
a standard nonlinear second--order Lagrangian field theory. The procedure entails a 
decomposition of
the dynamical data, consisting of the metric field $\omet_{\mu\nu}$ and its 
derivatives up to the third order,
into a set of independent fields describing the physical state by their
values and their first derivatives only. In this peculiar sense, one may say that 
the single ``unifying" field
$\omet_{\mu\nu}$ is replaced by (or decomposed into) a multiplet of gravitational fields.

An adequate mathematical tool for this purpose is provided by a
specific Legendre transformation \cite{MFF1, JK, MFF2}. Although the
transformation has been known for more than a decade, it is not currently
used in a systematic way. Instead, most papers on the nonlinear
spin--two field have employed various \emph{ad hoc} tricks adjusted to
quadratic Lagrangians \cite{ABJT, HOW1, Tom} (actually equivalent
to the Legendre transformation for this particular case), but such
approach does not allow one to fully exhibit the structure of the theory.

Although the Legendre transformation is essentially unique, the various fields 
of the resulting multiplet can be
given different physical interpretations; the different choices are traditionally 
called ``frames"\footnote{The use
of the word ``frame" in this sense should be deprecated, because it does not refer 
to the choice
of a physical reference frame, but this abuse of terminology is now so
universally adopted that we feel that trying to introduce here a more
appropriate term, for instance ``picture" as it is used in Quantum
Mechanics, would only lead to confusion.}. In general, the original
``Jordan frame" (JF; the name is borrowed from scalar--tensor theories)
consisting of only the unifying metric
$\omet_{\mu\nu}$, can be transformed into frames including fields of
definite spin in two ways. A first possibility is that the field $\omet_{\mu\nu}$
remains the spacetime metric, now carrying only two d.o.f., while the
other degrees of freedom (previously carried by its higher derivatives) are encoded
into auxiliary (massive) fields of definite spin: this is the Helmholtz--Jordan
frame (HJF). Alternatively, one introduces (via an appropriate
redefinition of the Legendre transformation) a new spacetime metric
$g_{\mu\nu}$, while the symmetric tensor $\omet_{\mu\nu}$ is decomposed into spin--2 and
spin--0 fields, forming in this way the massive, non--geometric components of the
gravitational multiplet; these variables form together the ``Einstein
frame" (EF). Both frames are dynamically equivalent and very similar on
the level of the field equations: the equations of motion are
second--order Lagrange field equations, and in each frame the
corresponding spacetime metric satisfies Einstein's field equations\footnote{Some
authors seem instead to believe
that Einstein equations  can be obtained only after redefinition of
the spacetime metric, i.e.÷in EF.}, thus
the theory looks like ordinary general relativity, with the
non--geometric components of the multiplet acting as specific matter
fields. The two frames differ however in the action integral. In HJF the spin--0 and spin--2 fields
are nonminimally coupled to gravity (to curvature) while there are no kinetic terms for those fields in the
Lagrangian; only the metric has the standard Einstein--Hilbert Lagrangian
$\oric(\omet)$. In consequence, propagation PDEs for the fields with spin zero and two
arise from the action in a more involved way (through the
variation of the metric tensor), so the theory in this frame cannot be
obtained by minimal coupling to ordinary gravity of some additional
fields already possessing a definite dynamics in a fixed
background spacetime. Yet it is remarkable that in the EF variables the
theory has fully standard form \cite{MFF1, JK}: one recovers the
$R(g)\sqrt{-g}$ Lagrangian for the metric and universal kinetic terms for
the spin--0 and spin--2 fields (independently of the form of the original
Lagrangian $L(\omet)$ in JF), only the potential part of the action
being affected by the actual form of $L(\omet)$. The EF variables are
uniquely characterised by these features, while in HJF different \emph{ad
hoc} redefinitions of the variables can be intertwined with the Legendre
transformation (the latter being itself sometimes disguised as a mere
change of variables) \cite{ABJT, HOW1}.

Though matematically equivalent, the two frames are physically
inequivalent; the difference is most clearly visible while defining the
energy since the latter is very sensitive to redefinitions of the
spacetime metric \cite{MS1}. Both mathematical similarity to ordinary
general relativity and physical arguments indicate that the EF is
physical \cite{MS1}: in HJF the energy--momentum tensor is unphysical,
being linear in both non--geometric fields, while in EF the
stress--energy tensor for the scalar field has the standard form and that
for the spin-two field seems also more acceptable than in HJF.

Anyhow, in both frames NLG theories provide a
consistent description of a self--gravitating massive spin--two field.
The field is necessarily nonlinear and in quantum theory it is ghostlike.
The latter defect is inferred from the fact that in the linearized theory
the Lagrangian of the field appears with the sign opposite to that for
linearized Einstein gravity \cite{St1}. This fact is interpreted in
classical theory as related to the occurrence of excitations with
negative energy for the field, and in consequence as a signal of
instability of the theory. However, in this paper we show (an
incomplete proof was given previously in \cite{HOW1}) that the ground
state solution (vacuum) \emph{is} classically stable at the linear level.
This does not prove the stability of the vacuum state whenever nonlinear
terms are taken into account: we stress, however, that all the
above mentioned features usually advocated as signals of instability, are
equally derived within the linear approximation. The problem of energy
is more subtle: in HJF the variational energy--momentum tensor is
evidently unphysical, while in EF the Lagrangian is highly nonlinear (not even polynomial), 
and we will show in sect.÷7 that the linear 
approximation tells rather little about the energy density of the exact theory. 
Hence, the massive spin--two field generated by NLG
theories is still worth investigating in the framework of classical
Lagrangian field theory.

Most previous works \cite{ABJT, HOW1, Tom} were centered on the particle
spectrum of the theory and dealt only with the action integrals,
while less attention was paid to the field equations and the structure of
the theory. The field equations in HJF were given in \cite{HOW1}, but
these authors regarded the field equations in EF as extremely involved
and thus intractable, they only studied the case where the spin--two
field is assumed to be proportional to the metric tensor and thus can be
described by a scalar function (its trace). In consequence, the dynamical
consistency of the theory has always been taken for granted since the
fourth--order equations in JF are consistent.

In the present paper we systematically investigate the nonlinear
spin--two field generated by a NLG theory with a quadratic Lagrangian
(\ref{n1}) in the framework of classical field theory, employing a
Legendre transformation. 
As is well known, a spin--two field may be mathematically represented by
tensor variables of different rank and symmetry properties \cite{AD1,
NN}. Any NLG theory generates in a natural way a representation of the
field in terms of a symmetric, second rank tensor
$\psi_{\mu\nu}=\psi_{\nu\mu}$, and this representation will be employed
in this work.

Although we study a concrete Lagrangian (this
is motivated in sects.~2 and 3) we employ no tricks adjusted to it. The
paper is self--contained and provides an (almost) full exposition of
the subject. The first part describes the nonlinear spin--two and
spin-zero fields in HJF. The particle content in this frame is well
known. The main new results shown here are: 
\begin{itemize}
\item the field equations for the
metric and the other two fields for any spacetime dimension $d\ge 4$; 
\item the fact that dimension $d=4$ is distinguished in that the scalar field is decoupled and can
be easily removed from the theory (we do so in the rest of the paper
since we are interested in the description of the spin--two field);
\item the fact (sect.~3) that the resulting
equations of motion for the spin--two field do not generate further
constraints besides the five ensuring the purely spin--2 character of the
field;
\item the internal consistency of the theory in HJF, ensured by
strong Noether conservation laws; 
\item the linear stability of the unique ground
state solution representing flat spacetime and vanishing spin--two field,
assessed by the fact that small perturbations form
plane waves with constant amplitudes. 
\end{itemize}
In sect.~4 we show that
the two possible ways to obtain the massless limit of the spin--two field
described in the previous section yield the same result, so the
massless limit is well--defined and leads to the propagation
equations for gravitational perturbations in a Ricci--flat spacetime.

The main thrust of the paper is its second
part (sect.÷5 to 9), where we investigate the equations of motion in Einstein frame.
Here most results are new. 
\begin{itemize}
\item A generic presentation of the Lagrangian theory for the nonlinear
spin--two field in Einstein frame is given in sect.~5: in order to
better exhibit the structure of the theory in this frame, we consider a
Lagrangian being an arbitrary function of the Ricci tensor of the
original metric in JF. We explicitly give the equations of motion for the
massive field in the generic case, an expression (highly nonlinear) for
the full energy--momentum tensor $T_{\mu\nu}$ of the field and four differential
constraints imposed on the field by its dynamics. 
\item This generic theory
is then specialized in sect.~6 to the case of the particular Lagrangian
of eq.~(\ref{n33}), which in HJF ensured that the scalar field drops out. The
previous, generic equations produce in this case a fifth, algebraic
constraint, which together with those already found ensures that also in
EF the massive field has five degrees of freedom and is
purely spin--two. 
\item A unique, linearly stable ground state solution is then
found (without any simplifying
assumptions); clearly it corresponds via
the Legendre transformation to the ground state in HJF. The spin--two field is then redefined to make it
vanish in this state (and in vacuum in general). 
It turns out that
the consistency and hyperbolicity problems in this frame, investigated in sect.~7, are harder that
in HJF and should be studied perturbatively; we study them in the
linearized theory.
The Lagrangian is computed in the lowest order
(quadratic) approximation around the ground state solution to show the
ghostlike character of the spin--two field. A detailed comparison is made with the theory of the linear spin--two
field, and the fact that the energy--momentum tensor of the nonlinear theory is not approximated by the
energy--momentum tensor derived from the Wentzel Lagrangian is fully explained.
\item Finally, contrary to the common belief that the
full (nonlinear) system of equations of motion in EF are intractably involved, in sect.~8 we give a simple but
nontrivial solution to them.
\end{itemize}

Conclusions are in sect.~9, and Lagrange field equations
and the energy--momentum tensor in EF for the redefined spin--two field
as well as some other useful formulae are contained in Appendix.

\section{Equations of motion for the gravitational \\ multiplet in 
Helmholtz--Jordan frame}
We will investigate dynamical structure and particle content of a 
nonlinear gravity (NLG) theory using Legendre transformation method 
\cite{MFF1, JK, MFF2}. The starting point is a $d$--dimensional manifold 
$M$, $d\geq 4$ (later the dimensionality will be fixed to $d=4$) endowed 
with a Lorentzian metric $\omet_{\mu\nu}$. The inverse contravariant 
metric tensor will be denoted by $\oimet^{\mu\nu}$ and $\oimet^{\mu\alpha}
\omet_{\alpha\nu}=\delta^{\mu}_{\nu}$; we introduce this nonstandard 
notation for further purposes. One need not view $\omet_{\mu\nu}$ as a 
physical spacetime metric, actually whether  $\omet_{\mu\nu}$ or its 
"canonically conjugate" momentum is the measurable quantity determining all 
spacetime distances in physical world should be determined only after a 
careful examination of the physical content of the theory, rather than 
prescribed {\it a priori}. Formally $\omet_{\mu\nu}$ plays
both the role of a metric  tensor on $M$ and is a kind of unifying field which will be decomposed in 
a multiplet of fields with definite spins; pure gravity is described in 
terms of the fields with the metric being a component of the multiplet. In 
general dynamics for $\omet_{\mu\nu}$ is generated by a nonlinear 
Lagrangian density $L\sqrt{-\omet}=f(\omet_{\mu\nu}, 
\oric_{\alpha\beta\mu\nu})\sqrt{-\omet}$ where $\omet\equiv \det(\omet_{\mu\nu})$  
and $\oric_{\alpha\beta\mu\nu}$ is the Riemann tensor for 
$\omet_{\mu\nu}$ ; $f$ is any smooth (not necessarily analytic) scalar 
function. Except for Hilbert--Einstein and Euler--Poincar\'e topological 
invariant densities the resulting variational Lagrange equations are of 
fourth order. The Legendre transformation technique allows one to deal 
with fully generic Lagrangians; from the physical standpoint, however, 
there is no need to investigate complicate or generic Lagrangians. Firstly, 
in the bosonic sector of low energy field theory limit of string effective 
action one gets in the lowest approximation the Hilbert--Einstein 
Lagrangian plus terms quadratic in the curvature tensor. Secondly, to 
obtain an explicit form of field equations and to deal with them 
effectively one needs to invert the appropriate Legendre transformation 
and in a generic case this amounts to solving nonlinear matrix equations. 
Hindawi, Ovrut and Waldram \cite{HOW2} have given arguments that a
generic NLG theory has eight degrees of freedom and the same particle
spectrum as in the quadratic Lagrangian (\ref{n1}) below, the only known physical
difference lies in the fact that in the generic case one expects multiple nontrival
(i.e.~different from flat spacetime) ground state solutions. This result can be also
derived from the observation that after the Legendre transformation the kinetic terms
in the resulting (Helmholtz) Lagrangian are universal, and only the potential terms
keep the trace of the original nonlinear Lagrangian. If the latter is a polynomial
of order higher than two in the curvature tensor, the Legendre map is only locally
invertible and this leads to multivalued potentials, generating a ground state
solution in each ``branch"; yet the form of the potential could produce additional
dynamical contraints, affecting the number of degrees of freedom, only in non--generic
cases. The physically relevant Lagrangians in field theory depend quadratically  on
generalized velocities and then conjugate momenta are linear functions of the
velocities. For both conceptual and practical purposes it is then  sufficient to
envisage a quadratic Lagrangian
\begin{equation}\label{n1}
L=\oric + a\oric^2 +b\oric_{\mu\nu}(\omet)
\oric^{\mu\nu}(\omet).
\end{equation}
In principle one should also include the term 
$\oric_{\alpha\beta\mu\nu}\oric^{\alpha\beta\mu\nu}$ (in 
four dimensions it can be eliminated via Gauss--Bonnet theorem), but the 
presence of Weyl tensor causes troubles: although formally the Legendre 
transformation formalism works well there are problems with providing 
appropriate propagation equations for the conjugate momentum and with 
physical interpretation (particle content) of the field. We therefore 
suppress Weyl tensor in the Lagrangian. The Lagrangian cannot be purely 
quadratic: it is known from the case of restricted NLG theories (Lagrangian 
depends solely on the curvature scalar, $L=f(\oric)$) that the 
linear term $\oric$ is essential \cite{MS1} and we will see that 
the same holds for Lagrangians explicitly depending on Ricci tensor 
$\oric_{\mu\nu}$. The coefficients $a$ and $b$ have dimension 
$\textrm{[length]}^2$; contrary to some claims in the literature there are no 
grounds to presume that they are of order $\textrm{(Planck length)}^2$ 
unless the Lagrangian (\ref{n1}) arises from a more fundamental theory (e.g. 
string theory) where $\hbar$ is explicitly present. Otherwise in a pure 
gravity theory the only fundamental constants are $c$ and $G$; then $a$ 
and $b$ need not be new fundamental constants, they are rather related 
to masses of the gravitational multiplet fields. Here we assume that the 
NLG theory with the Lagrangian (\ref{n1}) is an independent one, i.e. it inherits 
no features or relationships from a possible more fundamental theory. \\

As was mentioned in the Introduction, in the Legendre transformation one 
replaces the higher derivatives of the field 
$\omet_{\mu\nu}$ by additional fields.
In this section we assume that the original field $\omet_{\mu\nu}$ keeps the role of the physical 
spacetime metric, and the self--gravitating spin--two field originates from the
``conjugate momenta" to $\omet_{\mu\nu}$. \\

We recall that for a second--order Lagrangian such as (\ref{n1}) one should properly choose the 
quantities to be taken as generalized velocities to define, via a 
Legendre map, conjugate momenta \cite{MFF2}. One cannot, for instance, 
use the partial derivatives $\omet_{\mu\nu,\alpha\beta}$ as generalized velocities since for 
covariant Lagrangians, e.g. (\ref{n1}), the Legendre map cannot be inverted: the 
Hessian, being the determinant of a $100\times 100$ matrix, vanishes, 
\begin{equation}\label{n2}
\det \left(\frac{\partial^2 L}{\partial\omet_{\mu\nu,\alpha\beta}
\partial\omet_{\lambda\sigma,\rho\tau}}\right)=0.
\end{equation}
General covariance indicates which linear combinations of 
$\omet_{\mu\nu,\alpha\beta}$ can be used as the velocities, i.e. with 
respect to which combinations the Lagrangian is regular (the Hessian 
does not vanish). Clearly this is Ricci tensor $\oric_{\mu\nu}$. 
The explicit use of generally covariant quantities in this approach is 
supported by the Wald's theorem \cite{Wald1, CW} that only a generally 
covariant theory may be a consistent theory of a spin--two field. 
Following \cite{MFF2}, in order to decompose $\omet_{\mu\nu}$ into fields 
with definite spins, one makes Legendre transformations of the Lagrangian 
(\ref{n1}) with respect to the two irreducible components of 
$\oric_{\mu\nu}$: its trace $\oric$ and the traceless part 
$\osric_{\mu\nu}$. In terms of 
$\osric_{\mu\nu} \equiv \oric_{\mu\nu}-\frac{1}{d}
\oric\omet_{\mu\nu}$, 
$\osric_{\mu\nu}\oimet^{\mu\nu}=0$, the Lagrangian reads 
\begin{equation}\label{n3}
L= \oric +(a+\frac{b}{d})\oric^2+
b\osric_{\mu\nu}\osric^{\mu\nu},
\end{equation}
one assumes $ad+b\ne 0$ and $b \ne 0$. One then defines a scalar and a 
tensor canonical momentum via corresponding Legendre transformations:
\begin{equation}\label{n4}
\chi +1 \equiv \frac{\partial L}{\partial \oric}, \qquad 
\pi^{\mu\nu} \equiv \frac{\partial L}{\partial\osric_{\mu\nu}};
\end{equation}
it is convenient to identify $\partial L/\partial\oric$ with 
$\chi +1$ rather than with $\chi$ alone. From (\ref{n3}):
\begin{equation}\label{n5}
\chi = 2(a+\frac{b}{d})\oric \qquad \textrm{and} \qquad 
\pi^{\mu\nu}=2b\osric^{\mu\nu}
\end{equation}
hence fields $\chi$ and $\pi^{\mu\nu}$ are dimensionless and 
$\pi^{\mu\nu}$ is traceless, $\pi^{\mu\nu}\omet_{\mu\nu}=0$. 
The new triplet of field variables \{$\omet_{\mu\nu}, 
\chi, \pi^{\mu\nu}$\} defines the Helmholtz--Jordan Frame (HJF). 

Equations of motion for this frame arise as 
variational Lagrange equations from Helmholtz Lagrangian \cite{He, 
MFF1, JK, MFF2}. First one constructs the Hamiltonian 
\begin{equation}\label{n6}
H=\frac{\partial L}{\partial \oric}\oric +
\frac{\partial L}{\partial\osric_{\mu\nu}}\osric_{\mu\nu}
- L
\end{equation}
expressed in terms of $\omet_{\mu\nu}$ and the canonical momenta, 
it reads 
\begin{equation}\label{n7}
H=\frac {d}{4(ad+b)}\chi^2 + \frac{1}{4b}\pi^{\mu\nu}\pi_{\mu\nu}, 
\end{equation}
here $\pi_{\mu\nu}=\omet_{\mu\alpha}\omet_{\nu\beta}\pi^{\alpha\beta}$; 
all indices are raised and lowered with the aid of $\oimet^{\mu\nu}$ 
and $\omet_{\mu\nu}$. Next one evaluates Helmholtz Lagrangian 
defined as 
\begin{equation}\label{n8}
L_H \equiv \frac{\partial L}{\partial\oric}\oric
(\omet, \partial\omet, \partial^2\omet) +
\frac{\partial L}{\partial\osric_{\mu\nu}}\osric_{\mu\nu}
(\omet, \partial\omet, \partial^2\omet) - H(\omet, \chi, \pi), 
\end{equation}
where the derivatives $\partial L/\partial\oric$ and 
$\partial L/\partial\osric_{\mu\nu}$ are set equal to the canonical 
momenta $\chi +1$ and $\pi^{\mu\nu}$ respectively, while the "velocities" 
$\oric$ and $\osric_{\mu\nu}$ explicitly depend on first 
and second derivatives of $\omet_{\mu\nu}$. In classical mechanics 
for a first order Lagrangian $L(q, \dot{q})$ one has 
\begin{equation}\label{n9}
L_H (q, p, \dot{q}, \dot{p})\equiv p\dot{q}-H(q, p)= p\dot{q}-
p\dot{q}(q, p) - L(q, \dot{q}(q, p)), 
\end{equation} 
i.e. $L_H$ is a scalar function on the tangent bundle to the cotangent 
bundle to the configuration space; $L_H$ does not depend on $\dot{p}$. 
Similarly, in a field theory $L_H$ is independent of partial derivatives 
of canonical momenta. In classical mechnics the action 
$\int L_H \ud t$ gives rise, when varied with respect to $p$, to the 
equation $\dot{q}=\partial H/\partial p$, while varied with respect to 
$q$ generates 
\begin{equation}\label{n10}
\frac{\ud}{\ud t}\left(\frac{\partial L_H}{\partial \dot{q}}\right) -
\frac{\partial L_H}{\partial q} =0;
\end{equation} 
the latter equation is equivalent to 
\begin{equation}\label{n11}
\dot{p}=-\frac{\partial H}{\partial q} \qquad \textrm{and} \qquad 
\frac{\ud}{\ud t}\left(\frac{\partial L}{\partial \dot{q}}\right) -
\frac{\partial L}{\partial q} =0.
\end{equation} 
Thus $L_H$ simultaneously generates both Hamilton and Lagrange 
equations of motion. In the case of NLG theories one is interested in 
replacing the fourth order Lagrange equations by the equivalent 
second order Hamilton ones. For the Lagrangian (\ref{n1}) $L_H$ reads 
\begin{equation}\label{n12}
L_H= \oric + \chi\oric + \pi^{\mu\nu}
\osric_{\mu\nu} - \frac{d}{4(ad+b)}\chi^2 - 
\frac{1}{4b}\pi^{\mu\nu}\pi_{\mu\nu}. 
\end{equation}
One sees that the linear Hilbert--Einstein Lagrangian for the metric 
field is recovered. This means that Hamilton equations for 
$\omet_{\mu\nu}$ are not just second order ones of any kind but exactly 
Einstein field equations\footnote{We use units $8\pi G =c=1$, the 
signature is $(-+++)$. We use all the conventions of \cite{HE}.}
$\oein_{\mu\nu}=\Told_{\mu\nu}(\omet, \chi, \pi)$. 
The nonminimal 
coupling interaction terms $\chi\oric$ and 
$\pi^{\mu\nu}\osric_{\mu\nu}$ will cause that $\Told_{\mu\nu}$ will 
depend on second derivatives of $\chi$ and $\pi^{\mu\nu}$ and will 
contain Ricci tensor. Since Hilbert--Einstein Lagrangian for the 
metric field in general relativity is $L_g(\omet)=\frac{1}{2}
\oric$, then $L_H=2L_g +2L_f$, where $L_f$ is the Lagrangian for 
the non--geometric components of gravity, $\chi$ and $\pi^{\mu\nu}$. 
The energy--momentum tensor is then 
\begin{eqnarray}\label{n13}
-\frac{1}{2}\sqrt{-\omet}\ \Told_{\mu\nu}(\omet, \chi, \pi)& \equiv &
\frac{\delta}{\delta\oimet^{\mu\nu}}(\sqrt{-\omet}\ L_f)\\ & = &
\frac{1}{2}\frac{\delta}{\delta\oimet^{\mu\nu}}\left[\sqrt{-\omet}
\left(\chi\oric + \pi^{\alpha\beta}
\osric_{\alpha\beta} - \frac{d}{4(ad+b)}\chi^2 - 
\frac{1}{4b}\pi^{\alpha\beta}\pi_{\alpha\beta} \right)\right].\nonumber
\end{eqnarray}
Explicitly the equations $\frac{\delta}{\delta\oimet^{\mu\nu}}L_H= 0$ 
read
\begin{eqnarray}\label{n14}
\oein_{\mu\nu}(\omet) & = & \Told_{\mu\nu}(\omet, \chi, \pi) =
-\chi\oein_{\mu\nu}+\chi_{;\mu\nu}-\omet_{\mu\nu}\obox\chi+
\frac{1}{2}\oric_{\alpha\beta}\pi^{\alpha\beta}\omet_{\mu\nu}-\nonumber\\
& & -\frac{1}{2}\pi^{\alpha\beta}{}_{;\alpha\beta}\omet_{\mu\nu}+
\pi^{\alpha}{}_{(\mu;\nu)\alpha}-\frac{1}{2}\pi_{\mu\nu;\alpha}{}^{;\alpha}
-\frac{d}{8(ad+b)}\chi^2\omet_{\mu\nu}-\nonumber\\
& & -\frac{1}{8b}\pi^{\alpha\beta}\pi_{\alpha\beta}\omet_{\mu\nu}-
\frac{1}{2b}\pi_{\mu}{}^{\alpha}\pi_{\alpha\nu}-
\frac{1}{d}\oric\pi_{\mu\nu},
\end{eqnarray}
here $f_{;\alpha}$ denotes the covariant derivative with respect to 
$\omet_{\mu\nu}$ and $\obox f= \oimet^{\mu\nu}f_{;\mu\nu} =
f_{;\mu}{}^{;\mu}$. As remarked above $\Told_{\mu\nu}$ more resembles the 
stress tensor for the conformally invariant scalar field \cite{Pen, 
Bek} than that for ordinary matter. The equations of motion for 
$\chi$ and $\pi^{\mu\nu}$ are purely algebraic and clearly coincide 
with (\ref{n5}), 
\begin{equation}\label{n15}
\frac{\delta L_H}{\delta \chi}=0 \Rightarrow  \oric= 
\frac{d}{2(ad+b)}\chi, \qquad \frac{\delta L_H}{\delta\pi^{\mu\nu}}= 0 
\Rightarrow \osric_{\mu\nu}=\frac{1}{2b}\pi_{\mu\nu}.
\end{equation} 
These equations can be recast in the form of Einstein ones, 
\begin{equation}\label{n16}
\oein_{\mu\nu}(\omet)=\frac{1}{2b}\pi_{\mu\nu} -
\frac{d-2}{4(ad+b)}\chi\omet_{\mu\nu}.
\end{equation}
Comparison of eqs.~(\ref{n14}) and (\ref{n16}) shows that \emph{for solutions\/} 
there exists a simple linear expression for the stress tensor: 
\begin{equation}\label{n17}
\Told_{\mu\nu}(\omet, \chi, \pi)= \frac{1}{2b}\pi_{\mu\nu} -
 \frac{d-2}{4(ad+b)}\chi\omet_{\mu\nu}.
\end{equation}
This relationship allows one to derive differential propagation 
equations for $\chi$ and $\pi^{\mu\nu}$. Before doing it we simplify 
the expression (\ref{n14}) for $\Told_{\mu\nu}$ with the aid of (\ref{n16}) by replacing 
$\oric_{\mu\nu}$ by $\chi$ and $\pi^{\mu\nu}$ and making use of 
the Bianchi identity for $\oein_{\mu\nu}$. The latter provides 
a first order constraint on $\chi$ and $\pi^{\mu\nu}$, 
\begin{equation}\label{n18}
\pi^{\mu\nu}{}_{;\nu} = \frac{(d-2)b}{2(ad+b)}\chi^{,\mu},
\end{equation}
the constraint is already solved with respect to $\chi^{,\mu}$. Upon 
inserting (\ref{n16}) into the r.h.s. of (\ref{n14}) one gets 
\begin{eqnarray}\label{n19}
\Told_{\mu\nu}(\omet, \chi, \pi)& = &
\chi_{;\mu\nu}-\frac{1}{2}\obox\pi_{\mu\nu}
+\pi^{\alpha}{}_{(\mu;\nu)\alpha}-
\frac{1}{2b}\pi_{\mu}{}^{\alpha}\pi_{\alpha\nu}
-\frac{ad+2b}{2(ad+b)b}\chi\pi_{\mu\nu}\nonumber\\& & 
+\omet_{\mu\nu}\left(-\obox\chi
-\frac{1}{2}\pi^{\alpha\beta}{}_{;\alpha\beta}+
\frac{d-4}{8(ad+b)}\chi^2+\frac{1}{8b}\pi^{\alpha\beta}\pi_{\alpha\beta}
\right).
\end{eqnarray}
Equating the trace $\Told_{\mu\nu}\oimet^{\mu\nu}$ computed from (\ref{n17}) to 
the trace of (\ref{n19}) and applying (\ref{n18}) one arrives at a quasilinear 
equation of motion for $\chi$, 
\begin{equation}\label{n20}
[4(d-1)a+db]\obox\chi - \frac{d-4}{2d}\left (\frac{ad+b}{b}
\pi^{\alpha\beta}\pi_{\alpha\beta}+d\chi^2\right ) -(d-2)\chi =0.
\end{equation}
The field $\chi$ is self--interacting and is coupled to $\pi^{\mu\nu}$, 
the term $\pi^{\alpha\beta}\pi_{\alpha\beta}$ acts as a source for 
$\chi$. Eq.~(\ref{n20}) is a Klein--Gordon equation with a potential and an 
external source. The mass of $\chi$ is 
\begin{equation}\label{n21}
m^2_{\chi}=\frac{d-2}{4(d-1)a+db}
\end{equation}
and may be of both signs depending on the parameters. 

To derive a propagation equation for $\pi^{\mu\nu}$ one replaces 
derivatives of $\chi$ in (\ref{n19}) by derivatives of $\pi^{\mu\nu}$ with 
the help of (\ref{n18}). Then $\Told_{\mu\nu}$ depends on $\chi$ via terms 
$\chi\pi_{\mu\nu}$ and $\chi^2\omet_{\mu\nu}$; the latter is eliminated 
with the aid of eq.~(\ref{n20}) and the reappearing term $\obox
\chi$ is again removed employing (\ref{n18}). The resulting expression for 
$\Told_{\mu\nu}$, which contains terms $\chi\pi_{\mu\nu}$ and 
$\chi\omet_{\mu\nu}$, is set equal to the r.h.s. of eq.~(\ref{n17}), then the 
terms $\chi\omet_{\mu\nu}$ cancel each other and finally one arrives at 
the following equation of motion for $\pi_{\mu\nu}$, 
\begin{eqnarray}\label{n25}
\obox\pi_{\mu\nu}  
-\frac{4(ad+b)}{(d-2)b}\pi_{\alpha(\mu}{}^{;\alpha}{}_{;\nu)}
-2\pi^{\alpha}{}_{(\mu;\nu)\alpha}
+\frac{1}{b}\pi_{\mu\nu}
+\frac{1}{b}\pi_{\mu\alpha}\pi_{\nu}{}^{\alpha}+\nonumber\\
+\frac{ad+2b}{(ad+b)b}\chi\pi_{\mu\nu}
+\omet_{\mu\nu}
\left(
\frac{2(2a+b)}{(d-2)b}\pi^{\alpha\beta}{}_{;\alpha\beta}
-\frac{1}{bd}\pi^{\alpha\beta}\pi_{\alpha\beta}
\right)& = & 0.
\end{eqnarray}

The triplet of gravitational fields is described by a coupled system 
of eqs.~(\ref{n16}), (\ref{n20}) and (\ref{n25}) and the constraint (\ref{n18}). The equations (\ref{n20}) 
and (\ref{n25}) are quasilinear and contain interaction and self--interaction 
terms which cannot be removed for dimensions $d>4$. \\ 

Since for generic dimension the dynamics of the fields $\pi_{\mu\nu}$ and $\chi$ 
cannot be decoupled, one can obtain some information on the individual behaviour of each field by considering
particular solutions in which only one of the two fields is excited.
\begin{enumerate}
\item Let $\pi_{\mu\nu}=0$. Then eq.~(\ref{n25}) holds identically while the 
constraint (\ref{n18}) implies $\chi= \textrm{const}$ and eq.~(\ref{n20}) reduces to a 
quadratic equation, $(d-4)\chi^2 +2(d-2)\chi =0$. One solves it 
separately for $d>4$ and for $d=4$. 
\begin{itemize}
\item[a)] $d>4$. There are two solutions, $\chi=0$ and $\chi =
-2(d-2)/(d-4)$. From eq.~(\ref{n16}) they correspond to $\oric_{\mu\nu}
=0$ and $\oric_{\mu\nu}=-\frac{d-2}{(d-4)(ad+b)}\omet_{\mu\nu}$ 
i.e. Einstein space respectively. Therefore there exist two distinct 
ground state solutions: Minkowski space for $\chi=0$ and 
$d$--dimensional de Sitter space or anti--de Sitter  
for $\chi<0$ depending on the sign of $ad+b$. 
\item[b)] For $d=4$ the equation has only one solution $\chi=0$ 
corresponding to $\oric_{\mu\nu}=0$ and there is a unique 
ground state solution $\omet_{\mu\nu}=\eta_{\mu\nu}$ (Minkowski metric) 
and $\pi_{\mu\nu}=\chi=0$. 
\end{itemize}
\item Let $\chi=0$. We consider solutions for $d>4$. The scalar field 
equation (\ref{n20}) generates the algebraic equation 
$\pi^{\alpha\beta}\pi_{\alpha\beta}=0$, besides $\pi^{\mu\nu}{}_{;\nu}
=0$ arising from (\ref{n18}). Eq.~(\ref{n25}) gets reduced, upon employing
the two equations, to a linear equation 
\begin{equation}\label{n26}
\obox\pi_{\mu\nu} +\frac{1}{b}\pi_{\mu\nu} - 
2\oric_{\alpha(\mu\nu)\beta}\pi^{\alpha\beta} =0.
\end{equation}
This shows that the mass of the tensor field is $m^2_{\pi}=-1/b$. 
\end{enumerate} 

\subsection{The four--dimensional case}
Dimension four is clearly distinguished by the scalar field 
eq.~(\ref{n20}). Setting $d=4$, one finds two relevant properties:
\begin{enumerate}
\item[(i)] both the self--interaction and the source in eq.~(\ref{n20})
vanish and $\chi$ satisfies the linear Klein--Gordon equation 
\begin{equation}\label{n22}
2(3a+b)\obox\chi - \chi=0;
\end{equation} 
for $3a+b>0$ the mass is real, 
\begin{equation}\label{n23}
m^2_{\chi}=\frac{1}{2(3a+b)}>0. \end{equation}
We shall see below that this allows one to decouple completely the
propagation equations for the fields $\pi_{\mu\nu}$ and $\chi$;

\item[(ii)] furthermore, for $3a+b=0$ eq.~(\ref{n22}) admits only one solution, $\chi=0$, 
so for this particular choice of coefficients the scalar field disappears from the theory: we shall exploit this
fact in the next section to concentrate our investigation only on the spin--two field.
\end{enumerate}
We stress that for $d>4$ vanishing
of  the coefficient of $\obox\chi$ in eq.~(\ref{n20}) does not imply that
that the scalar drops out. In fact, the equation gives rise to an 
algebraic relationship between the scalar and tensor fields, 
\begin{equation}\label{n24}
\pi^{\alpha\beta}\pi_{\alpha\beta}= \frac{4d(d-1)}{(d-2)^2}\chi^2 +
\frac{8d(d-1)}{(d-2)(d-4)}\chi, 
\end{equation}
which replaces a propagation equation for $\chi$.

To prove the statements above, one starts from eq.~(\ref{n25}), and for $d=4$ one
eliminates $\chi$ from the interaction term applying eq.~(\ref{n22}): then $\obox\chi$ is 
replaced by $\pi^{\alpha\beta}{}_{;\alpha\beta}$ with the help of (\ref{n18}). 
Then the equations of motion for $\pi_{\mu\nu}$ read 
\begin{eqnarray}\label{n27}
\obox\pi_{\mu\nu}  
-\frac{2}{b}(4a+b)\pi_{\alpha(\mu}{}^{;\alpha}{}_{;\nu)}
-2\pi^{\alpha}{}_{(\mu;\nu)\alpha}
+\frac{1}{b}\pi_{\mu\nu}
+\frac{1}{b}\pi_{\mu\alpha}\pi_{\nu}{}^{\alpha}+\nonumber\\
+\frac{4}{b^2}(2a+b)(3a+b)\pi^{\alpha\beta}{}_{;\alpha\beta}\pi_{\mu\nu}
+\omet_{\mu\nu}
\left(
\frac{2a+b}{b}\pi^{\alpha\beta}{}_{;\alpha\beta}
-\frac{1}{4b}\pi^{\alpha\beta}\pi_{\alpha\beta}
\right)& = & 0.
\end{eqnarray}

The linear equation (\ref{n22}) for $\chi$ and quasilinear eq.~(\ref{n27}) for 
$\pi_{\mu\nu}$ are decoupled. Equations (\ref{n27}) are linearly dependent since 
the trace (w.r.t. $\oimet^{\mu\nu}$) vanishes identically, hence there are 
9 algebraically independent equations (they satisfy also differential 
identities, see below). \\

For the special solution $\chi=0$ the differential constraint reads
$\pi^{\mu\nu}{}_{;\nu}=0$. Then eq.~(\ref{n27}) reduces to 
\begin{equation}\label{n28}
\obox\pi_{\mu\nu} +\frac{1}{b}\pi_{\mu\nu} - 
2\oric_{\alpha(\mu\nu)\beta}\pi^{\alpha\beta} -
\frac{1}{4b}\pi^{\alpha\beta} \pi_{\alpha\beta}\omet_{\mu\nu} =0 
\end{equation}
and the mass of the field is the same as in $d>4$, i.e. $m^2_{\pi}
=-1/b$. The masses of the massive components of the gravitational 
triplet, $m^2_{\chi}=[2(3a+b)]^{-1}$ and $m^2_{\pi}=-b^{-1}$, agree with 
the values found in the linear approximation to the quadratic Lagrangian 
for spin--0 and spin--2 fields by \cite{St1, St2} and \cite{Tey}. The 
non--tachyon condition is then $b<0$ and $3a+b>0$ \cite{AEL}. 

This condition shows that the $\oric^2$ term in the Lagrangian 
(\ref{n1}) is essential. In fact, if this term is absent and the Lagrangian 
reads $L=\oric +b\oric_{\mu\nu}\oric^{\mu\nu}$, 
then (for dimensionality four) 
\begin{eqnarray}\label{n29}
\oein_{\mu\nu} & = & \Told_{\mu\nu}(\omet, \chi, \pi)= \frac{1}{2b}
(\pi_{\mu\nu} -\chi\omet_{\mu\nu}), \\\label{n30}
& & \obox\chi -\frac{1}{2b}\chi =0 
\end{eqnarray} 
and
\begin{equation}\label{n31}
\obox\pi_{\mu\nu} +\frac{1}{b}(1+\chi)\pi_{\mu\nu} - 
2\oric_{\alpha(\mu\nu)\beta}\pi^{\alpha\beta}+
\frac{1}{2b}\omet_{\mu\nu}\left(\chi
-\frac{1}{2}\pi^{\alpha\beta} \pi_{\alpha\beta}\right)
-4\chi_{;\mu\nu}=0.
\end{equation}
The two fields have masses $m^2_{\chi}=1/(2b)$ and $m^2_{\pi}=-1/b$ and 
one of them is necessarily a tachyon. It is worth noting that in the case
of  restricted NLG theories, i.e. $L=f(\oric)$, the $a\oric^2$ 
term is also essential in the Taylor expansion of the Lagrangian: for 
$a>0$ Minkowski space is stable as the ground state solution of the 
theory, while for $a<0$ it is classically unstable and for $a=0$ the 
solution may be unstable or stable \cite{MS1}. \\

Finally we show that all nine algebraically independent equations (\ref{n27}) 
are hyperbolic propagation equations for $\pi_{\mu\nu}$ and they 
contain no constraint equations. To this end one replaces  
$\pi^{\mu\nu}{}_{;\nu}$ by $\chi^{,\mu}$ with the aid of (\ref{n18}) and one 
arrives at the following equations: 
\begin{eqnarray}\label{n32}
\obox\pi_{\mu\nu}
+\frac{1}{b}\pi_{\mu\nu}
+\frac{1}{b}\pi_{\mu\alpha}\pi_{\nu}{}^{\alpha}
-\frac{1}{4b}\omet_{\mu\nu}\pi^{\alpha\beta}\pi_{\alpha\beta}
-2\oric_{\alpha(\mu}\pi_{\nu)}{}^{\alpha}
-2\oric_{\alpha(\mu\nu)\beta}\pi^{\alpha\beta}=\nonumber\\
=\frac{2a+b}{4a+b}\left[
4\chi_{;\mu\nu}-\frac{1}{2(3a+b)}\chi\omet_{\mu\nu}
-\frac{2}{b}\chi\pi_{\mu\nu}
\right].
\end{eqnarray}
In this form the hyperbolicity of all the equations is evident. 

\section{The spin--2 field in the gravitational doublet in 
Helmholtz--Jordan  frame}

We have seen in the previous section that the generic quadratic 
Lagrangian (\ref{n1}), subject only to the non--tachyon condition $b<0$ and 
$3a+b>0$, describes a triplet of gravitational fields, HJF=
$\{\omet_{\mu\nu}, \chi, \pi^{\mu\nu}\}$, where the nongeometric fields in 
the triplet represent (on the quantum--mechanical level) interacting 
particles with positive masses. It has long been known that the theory
(\ref{n1}) has 8 degrees of freedom \cite{St2,FT,BC,BD}.
As it was first found in the linear 
approximation \cite{St1} and then in the exact theory \cite{HOW1,ABJT},
these degrees of freedom are carried by a massless
spin--2 field (graviton, 2 degrees of  freedom), a massive spin--2
field (5 d.o.f.) and a massive scalar field. 

We are interested in the dynamics and physical properties of the massive 
spin--2 field. In this context, the scalar is undesirable and its
presence only  makes the system of the equations of motion more involved.
One can get rid of the unwanted scalar by a proper choice of the
coefficients in  the original Lagrangian. As mentioned previously, for
$3a+b=0$ eq.~(\ref{n22})  has only one trivial solution\footnote{That the scalar degree of freedom disappears in this
case was previously found in \cite{St1,ABJT}.}
$\chi=0$. We therefore restrict
our  further study to the special Lagrangian 
\begin{equation}\label{n33}
L=\oric +a(\oric^2 -3\oric_{\mu\nu}
\oric^{\mu\nu})
\end{equation}
and denote $m^2 \equiv (3a)^{-1}$ assuming $a>0$. As it was noticed in 
\cite{St1,HOW1} this Lagrangian can be neatly expressed in terms of Weyl 
tensor, 
\begin{equation}\label{n34}
L= \oric + \frac{1}{2m^2}(L_{GB} -
\tilde{C}_{\alpha\beta\mu\nu} \tilde{C}^{\alpha\beta\mu\nu}) 
\end{equation}
where  $L_{\textrm{GB}} = \oric_{\alpha\beta\mu\nu} 
\oric^{\alpha\beta\mu\nu} -4\oric_{\mu\nu} 
\oric^{\mu\nu} +\oric^2$, the Gauss--Bonnet term, is 
a total divergence in four dimensions. \\

One formally repeats the operations of the previous section and 
replaces the expressions (\ref{n15}) to (\ref{n19}) by 
\begin{equation}\label{n35}
\oric =6m^2\chi, \qquad \osric_{\mu\nu}= -\frac{m^2}{2}
\pi_{\mu\nu}, 
\end{equation}  
\begin{equation}\label{n36}
\oein_{\mu\nu}=\Told_{\mu\nu}(\omet, \chi, \pi)= -\frac{m^2}{2}
(\pi_{\mu\nu} +3\chi\omet_{\mu\nu}),
\end{equation}
$\pi^{\mu\nu}{}_{;\nu} = -3\chi^{,\mu}$ and 
\begin{eqnarray}\label{n37}
\Told_{\mu\nu}(\omet, \chi, \pi)& = &
\chi_{;\mu\nu}-\frac{1}{2}\obox\pi_{\mu\nu}
+\pi^{\alpha}{}_{(\mu;\nu)\alpha}+
\frac{m^2}{2}\pi_{\mu}{}^{\alpha}\pi_{\alpha\nu}
-m^2\chi\pi_{\mu\nu}-\nonumber\\& & 
-\omet_{\mu\nu}\left(\obox\chi
+\frac{1}{2}\pi^{\alpha\beta}{}_{;\alpha\beta}+
\frac{m^2}{8}\pi^{\alpha\beta}\pi_{\alpha\beta}
\right)
\end{eqnarray}
respectively. The scalar field is still present in these equations. 
However the trace of eq.~(\ref{n37}) is $\Told_{\mu\nu}\oimet^{\mu\nu}=0$ while 
the trace of eq.~(\ref{n36}) yields $\Told_{\mu\nu}\oimet^{\mu\nu}= -6m^2\chi$. 
Then $\chi=0$  and the scalar drops out from the theory. Although the 
scalar field vanishes one cannot, however, remove it from the 
Lagrangian in HJF unless one imposes the constraint 
$\oric=0$ already on the level of the initial Lagrangian in 
HF. It is more convenient to deal with Lagrangians containing no 
Lagrange multipliers and therefore the auxiliary nondynamic (i.e. 
having no physical degrees of freedom) scalar $\chi$ remains in the 
Helmholtz Lagrangian (\ref{n12}) which now reads  
\begin{equation}\label{n38}
L_H= \oric + \chi\oric + \pi^{\mu\nu}
\osric_{\mu\nu} - 3m^2\chi^2 + 
\frac{m^2}{4}\pi^{\mu\nu}\pi_{\mu\nu}. 
\end{equation} 
The system of field equations for the gravitational doublet 
HJF=$\{\omet_{\mu\nu}, \pi^{\mu\nu}\}$, having together seven degrees 
of freedom, consists of Einstein field equations,  
\begin{equation}\label{n39}
\oein_{\mu\nu}(\omet) =\Told_{\mu\nu}(\omet, \pi)= -\frac{m^2}{2}
\pi_{\mu\nu} 
\end{equation}
and quasilinear hyperbolic propagation equations for $\pi_{\mu\nu}$, 
\begin{eqnarray}\label{n40}
\obox\pi_{\mu\nu} -m^2\pi_{\mu\nu} - 
2\oric_{\alpha(\mu\nu)\beta}\pi^{\alpha\beta}+
\frac{m^2}{4}\omet_{\mu\nu}\pi^{\alpha\beta} \pi_{\alpha\beta}
=-m^2\pi_{\mu\nu}-2\Told_{\mu\nu}(\omet, \pi)=0.
\end{eqnarray}

It should be stressed that, as is clearly seen from the method of 
deriving eqs.~(\ref{n20}) and (\ref{n25}), the eqs.~(\ref{n39}) 
and (\ref{n40}) are not simply the 
variational equations $\delta L_H/\delta \oimet^{\mu\nu} =0$ and 
$\delta L_H/\delta \pi^{\mu\nu} =0$ with the substitutions $\chi=0$ 
and $\oric=0$. 

The field $\pi^{\mu\nu}$ satisfies 5 constraints, 
$\pi^{\mu\nu}\omet_{\mu\nu} =0$ and $\pi^{\mu\nu}{}_{;\nu}=0$. Notice 
that the field equations give rise to no further constraints. In fact, 
the trace of (\ref{n40}) and divergence of (\ref{n39}) vanish identically if the 
constraints hold. A possible source of a further constraint is 
divergence of eq.~(\ref{n40}). It may be shown by a direct calculation that 
if the equations (\ref{n39}) and (\ref{n40}) hold throughout the spacetime and if 
the five constraints are satisfied everywhere, then divergence of eq.~(\ref{n40}) vanishes identically. This confirms the previous result 
\cite{St1, St2, Tey, ABJT, HOW1, AEL} that this field has spin two without
any admixture  of lower  spin fields. \\

We now investigate the internal consistency of the theory based on 
eqs.~(\ref{n39}) and (\ref{n40}). It is well known \cite{AD1, AD2, BD}
that a linear  spin--two field $\Phi_{\mu\nu}$, massive or massless, has
inconsistent  dynamics in the presence of gravitation since in a curved
non-empty  spacetime the field loses the degrees of freedom it had in
flat  spacetime and the five conditions which ensured there its purely 
spin--2 character, $\Phi_{\mu\nu}\eta^{\mu\nu}=0=
\Phi^{\mu\nu}{}_{,\nu}$, are replaced by four differential constraints 
imposed on the field and Ricci tensor\footnote{For linear fields with
spins higher than 2 it was long believed
\cite{ChD} that there was no easy way to have physical fields on anything 
but Minkowski space; only recently a progress has been made for
massive fields \cite{CPD}.}.  Here we are dealing with the nonlinear
spin--2 field
$\pi_{\mu\nu}$  and one expects that this field is consistent. This
expectation follows  from the dynamical equivalence of the Helmholtz
Lagrangian (\ref{n38}) to the  purely metric Lagrangian (\ref{n33}) of
the fourth--order theory and the latter  one (as well as any other NLG
theory with a Lagrangian being any smooth  scalar function of the
curvature tensor) is known to be consistent.  However the spin--2 field
theory in HJF should be a self-contained  one and one should show its
consistency without invoking its equivalent  fourth--order version. \\
We first notice a difference in the structures of the theories for the 
linear and nonlinear spin--two fields. For the linear field 
$\Phi_{\mu\nu}$ in Minkowski space one first derives (quite involved) 
Lagrange field equations and then either by employing the gauge 
invariance (for the massless field) or by taking the trace and 
divergence of the field equations for the massive one, one derives the 
five constraints $\Phi_{\mu\nu}\eta^{\mu\nu}=0=
\Phi^{\mu\nu}{}_{,\nu}$. The method does not work in a generic curved 
spacetime \cite{AD1}. For the nonlinear field $\pi_{\mu\nu}$ the 
tracelessness is a direct consequence of the tracelessness of 
$\osric_{\mu\nu}$ and of the field eq.~(\ref{n15}), which now reads (in 
terms of $L_f=\frac{1}{2}(L_H-\oric$) rather than of $L_H$) 
\begin{equation}\label{n41}
\frac{\delta L_f}{\delta \pi^{\mu\nu}} \equiv E_{\mu\nu}(\pi) = 
\frac{1}{2}\osric_{\mu\nu} +\frac{m^2}{4}\pi_{\mu\nu} =0;
\end{equation} 
in the Helmholtz Lagrangian (\ref{n38}) it is not assumed that $\pi_{\mu\nu}$ 
has vanishing trace. After eliminating the scalar field one gets 
$\oric=0$ and the other four constraints $\pi^{\mu\nu}{}_{;\nu}=0$ 
and the algebraic equation for $\pi_{\mu\nu}$ reduces to 
\begin{equation}\label{n42}
E_{\mu\nu} = 
\frac{1}{2}(\oric_{\mu\nu} +\frac{m^2}{2}\pi_{\mu\nu}) =0, 
\end{equation}
while the Einstein field equations for $\omet_{\mu\nu}$ are 
\begin{equation}\label{n43}
\frac{\delta}{\delta \oimet^{\mu\nu}}\left(\frac{1}{2}\sqrt{-\omet}
L_H\right) =
\oein_{\mu\nu} -\Told_{\mu\nu}(\omet, \pi) =0, 
\end{equation}
where $\Told_{\mu\nu}$ is given by eq.~(\ref{n37}) for $\chi=0$. Hence the 
constraints ensuring that $\pi_{\mu\nu}$ has 5 degrees of freedom hold 
whenever the field equations hold. Clearly the system (\ref{n42})--(\ref{n43}) is 
equivalent to the system (\ref{n39})--(\ref{n40}) but the former is more convenient 
for studying the consistency of the equations. To this end one employs 
the coordinate invariance of the action integral \cite{AD2}. Under 
an infinitesimal coordinate transformation $x'^{\mu} =x^{\mu} +
\varepsilon^{\mu}(x)$, $\vert \varepsilon^{\mu}\vert \ll 1$, the metric 
and the spin-2 field vary as $\delta\oimet^{\mu\nu} =2\varepsilon
^{(\mu;\nu)}$ and 
\begin{equation}\label{n44}
\delta\pi^{\mu\nu} \equiv -\textrm{L}_{\varepsilon}\pi^{\mu\nu} =
-\pi^{\mu\nu}{}_{;\alpha}\varepsilon^{\alpha} +
\pi^{\alpha\nu}\varepsilon^{\mu}_{;\alpha} +  
\pi^{\mu\alpha}\varepsilon^{\nu}_{;\alpha},
\end{equation} 
here L is the Lie derivative. 
The action integral 
\begin{equation}\label{n45}
S_f =\int L_f \sqrt{-\omet}\ \ud^4x 
\end{equation} 
is generally covariant, therefore it is invariant under the 
transformation 
\begin{equation}\label{n46}
\delta_{\varepsilon}S_f =  
\int \varepsilon^{\mu}
\left[
\Told_{\mu\nu}{}^{;\nu}-F\chi_{,\mu}
-E_{\alpha\beta}\pi^{\alpha\beta}{}_{;\mu}
-2E_{\mu\alpha;\beta}\pi^{\alpha\beta}
-2E_{\mu\beta}\pi^{\alpha\beta}{}_{;\alpha}\right]\sqrt{-\omet}\ \ud^4x=0,
\end{equation}
where $F \equiv \delta L_f/\delta \chi = \frac{1}{2}\oric -
3m^2\chi$ and $\delta\chi = -\varepsilon^{\mu}\chi_{,\mu}$. Thus 
the coordinate invariance implies a strong Noether conservation 
law 
\begin{equation}\label{n47}
\Told_{\mu\nu}{}^{;\nu}-F\chi_{,\mu}
-E_{\alpha\beta}\pi^{\alpha\beta}{}_{;\mu}
-2E_{\mu\alpha;\beta}\pi^{\alpha\beta}
-2E_{\mu\beta}\pi^{\alpha\beta}{}_{;\alpha}=0.
\end{equation}

Now assume that the field equations $F=0$ and $E_{\mu\nu}=0$ hold. 
Then also $E_{\mu\alpha ;\beta}=0$ holds and the identity reduces to 
$\Told_{\mu\nu}{}^{;\nu}=0$. This is a consistency condition for Einstein 
field equations (\ref{n43}). Divergence of $\Told_{\mu\nu}$ should vanish due to 
the system of field equations and constraints without giving rise to 
further independent constraints. \\
By a direct calculation one proves the following proposition: 
if the field equations $\chi=0$ and (\ref{n42}) and the five constraints 
$\pi^{\mu\nu}\omet_{\mu\nu} =0 =\pi^{\mu\nu}{}_{;\nu}$ hold 
throughout the spacetime,  
then the stress tensor given by eq.~(\ref{n37}) is divergenceless,  
$\Told_{\mu\nu}{}^{;\nu} \equiv 0$. 
This shows that the system (\ref{n42})--(\ref{n43}) is consistent. \\

Owing to the tracelessness of $\pi^{\mu\nu}$ there is only one 
ground state solution for the system (\ref{n39})--(\ref{n40})
\cite{HOW1, HOW2} (i.e.~the spacetime is maximally symmetric and
$\pi^{\mu\nu}$ is covariantly constant). This is  Minkowski space,
$\omet_{\mu\nu} =\eta_{\mu\nu}$ and
$\pi_{\mu\nu} =0$. This state is linearly stable 
since small metric perturbations $\omet_{\mu\nu} =\eta_{\mu\nu} +
h_{\mu\nu}$ and excitations of $\pi_{\mu\nu}$ around $\pi_{\mu\nu} =0$ 
can be expanded into plane waves 
$h_{\mu\nu} =\pi_{\mu\nu} =p_{\mu\nu}\exp(ik_{\alpha}x^{\alpha})$ 
with a constant wave vector $k_{\alpha}$,  $k_{\alpha}k^{\alpha} = 
-m^2$ and a constant wave amplitude $p_{\mu\nu}$ satisfying 
$p_{\mu\nu}\eta^{\mu\nu}= 0 = p_{\mu\nu}k^{\nu}$. 

\section{Massless spin--two field in HJF} 
The massive spin--2 field $\pi_{\mu\nu}$ has a finite range with the 
length scale $m^{-1}$. According to the principle of physical 
continuity \cite{BD} as the mass tends to zero, the long-range force 
mediated by $\pi_{\mu\nu}$ should have a smooth limit and in this limit 
it should coincide with the strictly infinite-range theory. We 
therefore consider two cases: first the exactly massless theory 
resulting from the Helmholtz Lagrangian for $m=0$ and then the field 
equations of the previous section in the limit $m \to 0$. \\

After setting $m=0$ in eq.~(\ref{n38}) it is convenient to express the 
resulting Lagrangian in terms of Einstein tensor 
$\oein_{\mu\nu} = \osric_{\mu\nu} -\frac{1}{4}
 \oric\omet_{\mu\nu}$  and the trace of the spin--2 field, 
$\pi =\pi^{\mu\nu}\omet_{\mu\nu}$. One gets              
\begin{equation}\label{n48}
L_H= \oric + (\chi +\frac{1}{4}\pi)\oric + 
\pi^{\mu\nu}\oein_{\mu\nu}.  
\end{equation} 
Owing to the Bianchi identity this Lagrangian is invariant under the 
gauge transformation 
\begin{equation}\label{n49}
\pi^{\mu\nu} \to \pi'^{\mu\nu} = \pi^{\mu\nu} + 
\varepsilon^{\mu ;\nu} + \varepsilon^{\nu ;\mu} \qquad 
\textrm{and} \qquad \chi \to \chi' =\chi -\frac{1}{2}\varepsilon^{\alpha}
{}_{;\alpha} 
\end{equation} 
with arbitrary vector $\varepsilon^{\mu}$\footnote{Another possible
decomposition of $\omet_{\mu\nu}$ in this frame into fields with
definite spin, as is done in \cite{HOW1}, yields a different gauge
transformation not affecting the scalar field.}.  The
scalar
$\chi +\frac{1}{4}\pi$ is gauge invariant. We notice that  the scalar
$\chi$ cannot be removed already on the level of the  Lagrangian $L_H$
since it would break the gauge invariance. Moreover  the term $\chi
\oric$ in
$L_H$ is essential to obtain appropriate  field equations. Hence in this
approach the scalar cannot be  eliminated by using the first principles.
One can only set
$\chi=0$ in  a specific gauge. \\
The field equations are 
\begin{displaymath}
\frac{\delta L_H}{\delta \chi} =\oric=0 \qquad 
\textrm{and} \qquad \frac{\delta L_H}{\delta \pi^{\mu\nu}} = 
\oric_{\mu\nu} -\frac{1}{4}\oric\omet_{\mu\nu} =0, 
\end{displaymath} 
which imply $\oric_{\mu\nu} =0$. It is clear that unlike in the 
massive case now one cannot recover the original fourth-order 
Lagrangian (\ref{n1}) since the two fields are here unrelated to Ricci 
tensor. The fields $\chi$ and $\pi_{\mu\nu}$ do not interact with the 
metric $\omet_{\mu\nu}$ which acts as an empty--spacetime background 
metric. This suggests in turn that the two components of the 
gravitational triplet are test fields on the metric background, e.g. 
some excitations, and one may expect that they satisfy linear equations 
of motion. Variation of $L_H$ with respect to the metric yields 
Einstein field equations which are reduced in comparison to eq.~(\ref{n14}), 
\begin{equation}\label{n50}
\oein_{\mu\nu} = \Told_{\mu\nu}(\omet, \chi, \pi) =
\chi_{;\mu\nu}-\omet_{\mu\nu}\obox\chi-
\frac{1}{2}\obox\pi_{\mu\nu}
-\frac{1}{2}\pi^{\alpha\beta}{}_{;\alpha\beta}\omet_{\mu\nu}+
\pi^{\alpha}{}_{(\mu;\nu)\alpha}=0;
\end{equation}
the last equality follows from $\oric_{\mu\nu}=0$. The scalar 
field $\chi$ does not appear here in the combination 
$\chi +\frac{1}{4}\pi$ as in the Helmholtz Lagrangian since upon 
varying with respect to the metric $\oimet^{\mu\nu}$ one has 
$\delta\chi=0$ while $\delta \pi =\pi^{\mu\nu}\delta \omet_{\mu\nu}$ 
($\pi^{\mu\nu}$ is a fundamental field, i.e. is independent of the 
metric). \\

The harmonic gauge condition $\pi^{\mu\nu}{}_{;\nu}=0$ is most 
convenient and the field equations for $\chi$ and $\pi^{\mu\nu}$ 
simplify (upon employing $\oric_{\mu\nu}=0$) to 
\begin{equation}\label{n51}
\oein_{\mu\nu} =
\chi_{;\mu\nu}-\omet_{\mu\nu}\obox\chi-
\frac{1}{2}\obox\pi_{\mu\nu}
+\oric_{\alpha(\mu\nu)\beta}\pi^{\alpha\beta}=0;
\end{equation}
Their trace provides an equation for $\chi$, 
\begin{equation}\label{n52}
\obox(6\chi +\pi) =0.
\end{equation}
The harmonic gauge condition does not fix the gauge uniquely 
and the remaining gauge freedom is generated by any vector 
$\varepsilon_{\mu}$ satisfying the equation   
\begin{equation}\label{n53}
\obox\varepsilon_{\mu} +\varepsilon^{\alpha}{}_{;\alpha
\mu}=0.
\end{equation}
The scalar function $6\chi +\pi$ is not gauge invariant and under 
the transformation (\ref{n49}) varies as $6\chi' +\pi' = 6\chi +\pi -
\varepsilon^{\alpha}{}_{;\alpha}$. For the residual gauge freedom 
the scalar $\varepsilon^{\alpha}{}_{;\alpha}$ satisfies the equation 
$\obox\varepsilon^{\alpha}{}_{;\alpha}=0$ which follows from 
(\ref{n53}) upon taking its divergence. Since both $6\chi +\pi$ and 
$\varepsilon^{\alpha}{}_{;\alpha}$ are solutions to the scalar wave 
equation, one can choose $\varepsilon^{\mu}$ such that 
$\varepsilon^{\alpha}{}_{;\alpha} = 6\chi +\pi$. The residual gauge 
freedom allows then one to use a specific gauge wherein $6\chi +\pi =
0$. Replacing $\chi$ by $\pi$ one gets equations of motion for 
$\pi_{\mu\nu}$, 
\begin{equation}\label{n54}
\obox\pi_{\mu\nu}
-2\oric_{\alpha(\mu\nu)\beta}\pi^{\alpha\beta}
+\frac{1}{3}\pi_{;\mu\nu}
-\frac{1}{3}\omet_{\mu\nu}\obox\pi=0.
\end{equation}
The trace of these equations vanishes identically, hence there are 9 
algebraically independent equations. The scalar $\pi$ is still 
present in (\ref{n54}) and $\pi_{\mu\nu}$ is subject to four constraints, 
thus it represents 6 degrees of freedom. \\

In this gauge one can take a special solution where the nondynamic 
field $\chi$ is zero. Then $\pi=0$ too and the field $\pi_{\mu\nu}$ 
is subject to 
\begin{equation}\label{n55}
\obox\pi_{\mu\nu}
-2\oric_{\alpha(\mu\nu)\beta}\pi^{\alpha\beta}=0
\end{equation}
and the five constraints $\pi^{\mu\nu}{}_{;\nu}=0 =\pi$, hence it has 
a definite spin equal two. \\
It may be shown that eqs.~(\ref{n55}) do not generate further constraints 
since their divergence vanishes identically provided 
$\oric_{\mu\nu} =0$ and $\pi^{\mu\nu}{}_{;\nu}=0$. \\

Secondly, one takes the limit $m \to 0$ in the equations of motion 
(\ref{n39}) and (\ref{n40}). Since the scalar $\chi$ has already been 
eliminated, 
the resulting field equations, $\oric_{\mu\nu} =0$ and (\ref{n55}), 
represent the full set of solutions rather than a special class; the 
five constraints hold. \\

In both cases the linear spin--2 field $\pi^{\mu\nu}$ is dynamically 
equivalent to gravitational perturbations of the background 
empty-spacetime metric, $\oric_{\mu\nu}(\omet) =0$. If the 
perturbed metric is $\omet_{\mu\nu} +h_{\mu\nu}$, then on the level of
the field equations and in the  traceless--transversal gauge one can
identify 
$h_{\mu\nu}$  with $\pi_{\mu\nu}$. In this sense the theory for a
massless spin--2  field $\pi_{\mu\nu}$ resulting from the Lagrangian
(\ref{n33}) is trivial. Yet \cite{HOW1} claim that from the form of their
Lagrangian for the field, which is very similar to (\ref{n48}), one can
infer that the field is ghostlike. Since both Lagrangians in HJF have no
kinetic terms, this conclusion can be reliably derived only in Einstein
frame and this will be done in Sect.7.

\section{Equations of motion for the gravitational doublet in 
Einstein  frame: a generic Lagrangian} 
We have seen that the multiplet of fields forming HJF, in which the 
unifying field $\omet_{\mu\nu}$ was decomposed, is suitable for 
investigating the dynamical evolution, consistency and particle 
content of the theory. This, however, does not imply that the fields 
of this frame are truly physical variables. The problem of physical 
variables first appeared for a system of a scalar field interacting 
with gravity, since in this system one can make arbitrary 
redefinitions (see \cite{MS1} and references therein). Following 
\cite{MS1} we say that different formulations (employing different 
frames) of a theory provide various versions of the same theory if 
the frames are dynamically equivalent. Dynamical equivalence means that
Lagrange equations of motion in different frames are equivalent while
their action integral are, in general, unrelated; in classical field
theory this is sufficient to regard these frames as various
manifestations of one theory. Recently is has been shown that the two
most important frames, JF and EF, are equivalent not only in the above
classical sense: in the quantum context the path integral for the
Lagrangian (\ref{n1}) in JF is equal to the path integral in EF for
Einstein gravity coupled to a massive spin--2 field and a massive scalar
\cite{Tom}. In both frames many physical quantities are the same,
e.g.~for black holes all the thermodynamical dynamical variables do not
alter under a suitable Legendre map \cite{KoMa} and the Zeroth Law and
the Second Law of black hole thermodynamics are proved in EF \cite{JKM}
(it is worth noticing that the proof works provided the coefficient of
the $R^2$ term in the Lagrangian in JF is positive). It is not quite
clear whether the latter proof and many other computations mentioned in
\cite{MS1} mean that EF is the physical frame or merely show that it is
computationally advantageous. The physical frame is 
distinguished among all possible dynamically equivalent ones in a rather
subtle way: its field variables are operationally measurable and field
excitations above the stable ground state solution have  positive energy
density (or satisfy the dominant energy condition).  In the case of a
scalar field appearing in scalar--tensor gravity  theories or arising in
the restricted NLG theories this criterion  uniquely points to the
physical variables (frame): the physical spacetime metric is conformally
related (by a degenerate Legendre transformation) to the metric field of
Jordan frame in which the theory has been originally formulated
\cite{MS1}, and coincides with the EF metric. We emphasize that energy
density is very sensitive to field redefinitions and thus is a good
indicator of which variables are physical. In this sense  the
energy--momentum tensor for the spin--2 field, eq.~(\ref{n39}), being 
just proportional to
$\pi_{\mu\nu}$, is unphysical\footnote{Also a different definition of a
tensor $\pi_{\mu\nu}$, representing the spin--two field in HJF, given
in \cite{HOW1}, results in a linear energy--momentum tensor.}. In fact,
in  absence of any empirical evidence regarding energy density for spin--2
particles, one may invoke analogy with a scalar field. \\

Though a scalar field also has not been observed yet, it is generally 
accepted that its energy--momentum tensor should be purely quadratic 
in the field derivatives and no linear terms may appear (in the kinetic 
part). For any long--range scalar field the presence of such linear 
terms would cause difficulties in determining its total energy. 
Given any test scalar field $\Phi$ in an empty spacetime 
($R_{\mu\nu}(g)=0$) one can take the tensor $\theta_{\mu\nu}= 
\Phi_{;\mu\nu} -g_{\mu\nu}\Box \Phi$, which is trivially conserved 
and dominates in the effective energy--momentum tensor at large 
distances and hence affects the total energy \cite{SGLA}. The 
degenerate Legendre transformation to the Einstein frame in 
scalar--tensor gravity and restricted NLG theories removes all 
linear terms and provides a fully acceptable expression for the 
scalar field energy density. \\

We therefore study now the other, more sophisticated way of 
decomposing $\omet_{\mu\nu}$ into a multiplet of fields by means of 
a Legendre transformation. We regard this transformation as a 
transition to the physical frame. We shall see that this 
transformation does not guarantee that the spin--2 field has 
positive energy density. Maybe the criterion for physical variables 
should be relaxed in the case of this field. In any case the new 
variables seem to be more physical than those in HJF. \\
One generates the physical spacetime metric $g_{\mu\nu}$ from 
$\omet_{\mu\nu}$ while the latter is viewed as a non-geometric 
component of gravity. It is useful and instructive to take at the 
outset the fully generic Lagrangian $L=f(\omet_{\mu\nu}, 
\oric_{\alpha\beta})$ and only later to specify it to the 
form (\ref{n33}). According to \cite{MFF1} and \cite{JK} the true 
spacetime metric is defined as\footnote{In \cite{ABJT}, \cite{HOW1} and
\cite{Tom} the physical metric is constructed in a more involved and
tricky way; the outcome is equivalent to the Legendre map (\ref{n56}).} 
\begin{equation}\label{n56}
g^{\mu\nu}\equiv
(-\omet)^{-1/2}\left|\det\left(\frac{\partial
f}{\partial\oric_{\alpha\beta}}\right)\right|^{-1/2}
\frac{\partial
f}{\partial\oric_{\mu\nu}}=\left|\frac{\omet}{g}\right|^{1/2}
\frac{\partial
f}{\partial\oric_{\mu\nu}},
\end{equation}
where $\omet =\det(\omet_{\mu\nu})$, $g =\det(g_{\mu\nu})$ and 
$g_{\mu\nu}$ is the inverse of $g^{\mu\nu}$, $g^{\mu\alpha}
g_{\alpha\nu} =\delta^{\mu}_{\nu}$. To view $g^{\mu\nu}$ as a 
spacetime metric one assumes that $\det(\partial f/\partial
\oric_{\alpha\beta}) \ne 0$. From now on all tensor 
indices will be raised and lowered with the aid of this metric.
At this point, to make the following equations more readable, we alter
our notation and denote the original tensor field $\omet_{\mu\nu}$ by
$\psi_{\mu\nu}$ and its inverse $\oimet^{\mu\nu}$ by
$\gamma^{\mu\nu}$.
\renewcommand\omet{\psi}
\renewcommand\oimet{\gamma}
The Legendre transformation (\ref{n56}) is a map of the metric 
manifold $(M, \omet_{\mu\nu})$ to another one, $(M, g_{\mu\nu})$. 
We will not consider here the hard problem of whether the 
transformation is globally invertible\footnote{Some basic
considerations of the problem can be found in \cite{HOW1}.}; we assume
that the map  is regular in some neighbourhood of a ground state
solution.  The fields
$g_{\mu\nu}$ and
$\omet_{\mu\nu}$ will be referred to as  Einstein  frame, EF
=$\{g_{\mu\nu},
\omet_{\mu\nu}\}$.  For the generic Lagrangian $\omet_{\mu\nu}$ is
actually a mixture of  fields carrying spin two and zero. Notice that for
$f$ as in (\ref{n1}), 
\begin{equation}\label{n57}
g^{\mu\nu} =\left\vert\frac{\omet}{g}\right\vert^{1/2}[(1+ 
2a\oric)\oimet^{\mu\nu} +2b \oric^{\mu\nu}], 
\end{equation}
hence for $\omet_{\mu\nu}$ being Lorentzian and close to Minkowski 
metric, $g_{\mu\nu}$ is also Lorentzian and close to flat metric 
(and thus invertible). This shows the importance of the linear term 
in (\ref{n1}). Furthermore, in the limit $m_{\pi} \to \infty$, i.e. 
$b \to 0$, the spin--2 field $\pi_{\mu\nu}$ becomes a non-dynamic 
one since it is determined in terms of $\chi$ and $\omet_{\mu\nu}$ 
by an algebraic equation, 
\begin{equation}\label{n58}
(1+\chi)\pi_{\mu\nu} +\pi_{\mu\alpha} \pi_{\nu}{}^{\alpha} -
\frac{1}{4}\omet_{\mu\nu} \pi^{\alpha\beta}\pi_{\alpha\beta} =0.
\end{equation}
($\pi_{\mu\nu}=0$ in the linear approximation.) Then (\ref{n57}) is 
reduced to the conformal rescaling of the metric \cite{HBW} (for 
more references cf.~\cite{MS1}) being the degenerate Legendre 
transformation \cite{MFF1, JK}, 
\begin{equation}\label{n59}
g_{\mu\nu} =\left(\lim_{b \to 0}\frac{\partial f}{\partial 
\oric}\right)\omet_{\mu\nu} = (1 +2a\oric)\omet_{\mu\nu}. 
\end{equation} 
As mentioned above, for a restricted NLG theory, 
$L=f(\oric)$, Einstein frame is the physical one.  \\

For the time being we investigate the generic Lagrangian. We shall 
use a tensor being the difference of the two Christoffel 
connections, \cite{MFF1, HOW1}, 
\begin{equation}\label{n60}
Q^{\alpha}_{\mu\nu}(g, \omet)\equiv\tilde\Gamma^{\alpha}_{\mu\nu}(\omet)-
\Gamma^{\alpha}_{\mu\nu}(g) = \frac{1}{2}\oimet^{\alpha\beta}
\left(\omet_{\beta\mu;\nu}+\omet_{\beta\nu;\mu}-\omet_{\mu\nu;\beta}\right),
\end{equation}
hereafter $\nabla_{\alpha} T\equiv T_{;\alpha}$ denotes the covariant 
derivative of any $T$ with respect to the physical metric $g_{\mu\nu}$. 
For any two metric tensors (not necessarily related by a 
transformation) the following identity holds for their curvatures 
\cite{MFF1}, 
\begin{eqnarray}\label{n61}
K^{\alpha}{}_{\beta\mu\nu}(Q) & \equiv &
\oric^{\alpha}{}_{\beta\mu\nu}(\omet)
-R^{\alpha}{}_{\beta\mu\nu}(g)=\nonumber\\& = &
Q^{\alpha}_{\beta\nu;\mu}
-Q^{\alpha}_{\beta\mu;\nu}
+Q^{\sigma}_{\beta\nu}Q^{\alpha}_{\sigma\mu}
-Q^{\sigma}_{\beta\mu}Q^{\alpha}_{\sigma\nu} =
-K^{\alpha}{}_{\beta\nu\mu}.
\end{eqnarray}
This ``curvature difference tensor" for $Q$ satisfies  
$K^{\alpha}{}_{\alpha\mu\nu}(\omet,g) =0$ and generates a 
``Ricci difference tensor"
\begin{equation}\label{n62}
K_{\mu\nu}(Q) \equiv
\oric_{\mu\nu}(\omet)
-R_{\mu\nu}(g)= 
Q^{\alpha}_{\mu\nu;\alpha}
-Q^{\alpha}_{\alpha\mu;\nu}
+Q^{\alpha}_{\mu\nu}Q^{\beta}_{\alpha\beta}
-Q^{\alpha}_{\mu\beta}Q^{\beta}_{\nu\alpha} = K_{\nu\mu}
\end{equation}
since $Q^{\alpha}_{\alpha\mu ;\nu} = Q^{\alpha}_{\alpha\nu ;\mu}$, 
and a ``curvature difference scalar" 
\begin{equation}\label{n63}
g^{\mu\nu}K_{\mu\nu}(\omet,g) = 
\nabla_{\alpha}\left(g^{\mu\nu}Q^{\alpha}_{\mu\nu}
-g^{\alpha\mu}Q^{\beta}_{\beta\mu}\right)
+g^{\mu\nu}\left(Q^{\alpha}_{\mu\nu}Q^{\beta}_{\alpha\beta}
-Q^{\alpha}_{\mu\beta}Q^{\beta}_{\nu\alpha}\right).
\end{equation}

Second order field equations in EF are generated by a Helmholtz 
Lagrangian \cite{MFF1,JK}. First one inverts the relationship (\ref{n56}), 
i.e. solves it with respect to Ricci tensor, 
$\oric_{\mu\nu}(\omet_{\alpha\beta}, \partial\omet_{\alpha\beta}, 
\partial^2\omet_{\alpha\beta}) =r_{\mu\nu}(g^{\alpha\beta}, 
\omet_{\alpha\beta})$. The functions $r_{\mu\nu}$ do not contain 
derivatives of $g^{\alpha\beta}$ and $\omet_{\alpha\beta}$. A unique 
solution exists providing that the Hessian 
\begin{displaymath}
\det\left(\frac{\partial^2 f}{\partial \oric_{\alpha\beta}
\oric_{\mu\nu}}\right) \ne 0.
\end{displaymath}
This means that $f$ must explicitly depend (at least quadratically) 
on $\oric_{\mu\nu}$ and not only on $\oric$. It is here 
that the assumption that the original Lagrangian is at most 
quadratic is of practical importance. Next one constructs the 
Hamiltonian density (it is more convenient to use at this point scalar 
densities than pure scalars)
\begin{equation}\label{n64}
H(g, \omet) \equiv g^{\mu\nu} r_{\mu\nu}\sqrt{-g} -
f\left(\omet_{\mu\nu},  r_{\alpha\beta}(g, \omet)\right)\sqrt{-\omet}
\end{equation}
and then a Helmholtz Lagrangian density, 
\begin{equation}\label{n65}
L_H\sqrt{-g} \equiv g^{\mu\nu}\oric_{\mu\nu}(\omet)\sqrt{-g} -
H(g, \omet).
\end{equation}
From (\ref{n62}) one finds $\oric_{\mu\nu}(\omet)= 
R_{\mu\nu}(g) + K_{\mu\nu}(Q)$, then 
\begin{equation}\label{n66}
L_H(g, \omet)= R(g) + g^{\mu\nu}K_{\mu\nu}(Q) -
g^{\mu\nu}r_{\mu\nu}(g, \omet) +
\left\vert\frac{\omet}{g}\right\vert^{1/2}
f\left(\omet_{\mu\nu},  r_{\alpha\beta}(g, \omet)\right).
\end{equation}

It is remarkable \cite{MFF1, JK} that in EF the Hilbert--Einstein 
Lagrangian for the spacetime metric is recovered and the kinetic part 
of the Lagrangian for $\omet_{\mu\nu}$ is universal, i.e. is 
independent of the choice of the scalar function $f$. The only 
reminiscence of the original Lagrangian $L$ in JF is contained in 
the potential part of $L_H$. It was far from being obvious that it is 
possible to define the physical metric in such a way that the 
gravitational part of $L_H$ is exactly $R(g)$. In this sense Einstein 
general relativity is a universal Hamiltonian image (under the 
Legendre map) of any NLG theory. The total divergence appearing in 
eq.~(\ref{n63}) may be discarded and the kinetic Lagrangian for 
$\omet_{\mu\nu}$ reads 
\begin{equation}\label{n67}
K(Q) \equiv g^{\mu\nu} (Q^{\alpha}_{\mu\nu}Q^{\beta}_{\alpha\beta} -
Q^{\alpha}_{\mu\beta}Q^{\beta}_{\nu\alpha}), 
\end{equation}
what ensures that Lagrange equations of motion will be of second 
order. The proof that $L_H$ in EF is dynamically equivalent to $L$ 
in JF is given in \cite{JK} and \cite{MFF1}. As in HJF we write 
$L_H=2L_g +2L_M$ with 
\begin{equation}\label{n68}
L_M =\frac{1}{2}K -\frac{1}{2}g^{\mu\nu}r_{\mu\nu} +
\frac{1}{2}\left\vert\frac{\omet}{g}\right\vert^{1/2}
f(\omet_{\alpha\beta}, r_{\alpha\beta}(g, \omet))
\end{equation} 
being the Lagrangian for $\omet_{\mu\nu}$. This Lagrangian was found in a different
way in \cite{HOW1} and \cite{Tom} for a different decomposition of the original
metric $\omet_{\mu\nu}$.\\

Equations of motion for the metric are Einstein ones, $G_{\mu\nu}(g)= 
T_{\mu\nu}(g, \omet)$ where as usual 
\begin{displaymath}
T_{\mu\nu}=-\frac{2}{\sqrt{-g}}\frac{\delta}{\delta g^{\mu\nu}}
(\sqrt{-g}L_M).
\end{displaymath}
However finding out $T_{\mu\nu}$ directly from $L_M$ requires a very 
long computation. To compute the dependence of $T_{\mu\nu}$ on the
dynamical variables for solutions one can use an alternative
derivation. The fundamental differential relation between the two
metrics, 
$\oric_{\mu\nu}(\omet_{\alpha\beta}) = 
r_{\mu\nu}(g^{\alpha\beta}, \omet_{\alpha\beta})$, which represents the inverse
Legendre map and is recovered in EF as one of the Euler--Lagrange equations, is inserted in
eq.(\ref{n62}) giving rise to
$R_{\mu\nu}(g)= r_{\mu\nu} - K_{\mu\nu}$. Next, one takes into account
that for solutions the variational energy--momentum tensor (or stress
tensor) $T_{\mu\nu}$ should be equal to the Einstein tensor, 
\begin{equation}\label{n69}
T_{\mu\nu}(g, \omet) = G_{\mu\nu}(g) = -K_{\mu\nu}(Q) +
\frac{1}{2}g_{\mu\nu}g^{\alpha\beta}
\left(K_{\alpha\beta}-r_{\alpha\beta}\right)+r_{\mu\nu}.
\end{equation} 
The second equality becomes an identity upon insertion of the Legendre map. It may be
verified that this expression coincides with that arising from the definition of the
stress tensor, hence it is valid not only for solutions. As a consequence, the
Einstein field equations turn into an identity when $g_{\mu\nu}$ and $\omet_{\mu\nu}$
are related by the Legendre transformation (\ref{n56}). From the Lagrange equations
of motion for $\omet_{\mu\nu}$ (see below) it will be evident that the use of the
equations does not simplify the formula for $T_{\mu\nu}$.
Since  $Q^{\alpha}_{\mu\nu}$ contains
$\oimet^{\mu\nu}$, the  energy--momentum tensor is a highly nonlinear function of
$\omet_{\mu\nu}$;  in general it also comprises linear terms. The presence of second 
order derivatives of $\omet_{\mu\nu}$ in $T_{\mu\nu}$ means that the 
energy density is not determined by initial data on a Cauchy surface. 
This is a generic feature of the stress tensors for vector and tensor fields: the
vector gauge fields (represented by one--forms) and scalar fields are the only
exceptions.
\\

The two metrics are subject to Bianchi identities. That for $g_{\mu\nu}$, 
$G^{\mu\nu}{}_{;\nu}=0$, is of less practical use while the other one, 
$\onab_{\nu}\oein^{\nu}_{\mu}(\omet)=0$, may be 
reformulated by setting $\oric_{\mu\nu}=r_{\mu\nu}(g, \omet)$ 
and replacing $\onab_{\nu}$ by $\nabla_{\nu}$ according to 
$\onab_{\nu}A^{\alpha} = \nabla_{\nu}A^{\alpha} +
Q^{\alpha}_{\nu\sigma}A^{\sigma}$ for any vector field. In this way 
the identity is transformed into first order equations for $r_{\mu\nu}$, 
\begin{equation}\label{n70}
\oimet^{\alpha\beta}(r_{\mu\alpha ;\beta} -\frac{1}{2}
r_{\alpha\beta ;\mu} -  Q^{\sigma}_{\alpha\beta}r_{\mu\sigma}) =0.
\end{equation}
These constitute four differential constraints on $\omet_{\mu\nu}$; they 
are independent of the equations of motion for the field. \\

The Lagrange equations of motion are derived in the standard way in a 
simple but long and strenuous computation. These read 
\begin{eqnarray}\label{n71}
E^{\mu\nu} & \equiv & \nabla_{\alpha}\left(\frac{\partial (2 L_M)}
{\partial \omet_{\mu\nu;\alpha}}\right) - \frac{\partial (2 L_M)}
{\partial \omet_{\mu\nu}} =
\oimet^{\alpha\beta}\left(K^{(\mu}{}_{\alpha\beta}{}^{\nu)}  +
Q^{(\mu\ ;\nu)}_{\alpha\beta}
-\frac{1}{2}g^{\mu\nu}Q^{\sigma}_{\sigma\alpha;\beta} \right)\nonumber\\
&& -\oimet^{\alpha(\mu}g^{\nu)\beta}K_{\alpha\beta} - \oimet^{\alpha(\mu}Q^{\nu)\
;\beta}_{\alpha\beta}
+\frac{1}{2}\oimet^{\mu\nu}g^{\alpha\beta}Q^{\sigma}_{\alpha\beta;\sigma}
-2\oimet^{\alpha\beta}Q^{\sigma}_{\sigma\alpha}Q^{(\mu}_{\beta\tau}g^{\nu)\tau}\nonumber\\
&&
+g^{\alpha\beta}\left(Q^{\sigma}_{\sigma\tau}Q^{(\mu}_{\alpha\beta}\oimet^{\nu)\tau}
-Q^{\sigma}_{\alpha\tau}Q^{(\mu}_{\sigma\beta}\oimet^{\nu)\tau}
+Q^{(\mu}_{\alpha\sigma}Q^{\nu)}_{\tau\beta}\oimet^{\sigma\tau}\right)\nonumber\\
&&
+\frac{1}{2}g^{\mu\nu}Q^{\beta}_{\alpha\beta}
\left(\oimet^{\alpha\tau}Q^{\sigma}_{\sigma\tau}
+\oimet^{\sigma\tau}Q^{\alpha}_{\sigma\tau}\right)-P^{\mu\nu}\equiv
M^{\mu\nu}(\nabla\omet) - P^{\mu\nu} = 0,
\end{eqnarray}
where $M^{\mu\nu}(\nabla\omet)$ denotes the kinetic part of the equations 
comprising all derivative terms and $P^{\mu\nu}$ is the potential 
part, 
\begin{equation}\label{n72}
P^{\mu\nu} \equiv \frac{\partial}{\partial \omet_{\mu\nu}}
\left[-g^{\alpha\beta}r_{\alpha\beta}(g, \omet) + 
\left\vert\frac{\omet}{g}\right\vert^{1/2}
f(\omet_{\alpha\beta}, r_{\alpha\beta}(g, \omet))\right].
\end{equation} 
The kinetic part is \emph{universal} while $P^{\mu\nu}$ explicitly 
depends on $f$ and $r_{\alpha\beta}$. At first sight these equations 
seem intractably complicated but we shall see that for the special 
Lagrangian (\ref{n33}) they may be simplified and some interesting 
information can be extracted as well as a simple solution
can be found. \\

Since in general $\omet_{\mu\nu}$ is a mixture of spin--0 and spin--2 
fields, there are only four constraints (\ref{n70}). Only after extracting 
and removing the scalar field one can derive a fifth constraint. In 
the case of the linear massive spin--2 field in flat spacetime the 
fifth constraint is generated by the trace of the equations of 
motion \cite{AD1} and something analogous occurs in the present case 
though the procedure is more involved. The trace $E^{\mu\nu}
g_{\mu\nu}=0$ is a second order equation while that with respect to 
$\omet_{\mu\nu}$ is (after many manipulations) 
\begin{equation}\label{n73}
E^{\mu\nu}\omet_{\mu\nu} = \oimet^{\mu\nu}(J_{\mu ;\nu}- 
J_{\mu}\omet_{\nu\alpha ;\beta}\oimet^{\alpha\beta}) -
P^{\mu\nu}\omet_{\mu\nu} = 0, 
\end{equation}
where $J_{\mu} \equiv \omet_{\mu\alpha}{}^{;\alpha} -\frac{1}{4}\tau
\oimet^{\alpha\beta}\omet_{\alpha\beta ;\mu}$ and 
$\tau \equiv g^{\mu\nu}\omet_{\mu\nu}$. We will see in the next section 
that $J_{\mu}$ vanishes in the particular case (\ref{n33}), inducing an additional constraint. 

\section{Constraints and the ground state solution in EF}

From now on we restrict our study to the special quadratic Lagrangian 
(\ref{n33}) and expect that $\omet_{\mu\nu}$ will turn out to be a massive 
purely spin--2 non-geometric component of the gravitational doublet. 
The Legendre transformation (\ref{n57}) takes the form 
\begin{equation}\label{n74}
g^{\mu\nu}= A[(1 +\frac{2}{3m^2}\oric)\oimet^{\mu\nu} -
\frac{2}{m^2}\oric^{\mu\nu}]
\end{equation}
with 
\begin{displaymath}
A \equiv \left\vert\frac{\omet}{g}\right\vert^{1/2} =
\vert \det(\omet^{\alpha}_{\beta})\vert^{1/2}.
\end{displaymath}
Inverting this transformation one gets 
\begin{equation}\label{n75}
\oric_{\mu\nu}(\omet) =r_{\mu\nu}(g, \omet) = \frac{m^2}{2A}
[(\tau -3A)\omet_{\mu\nu} -\omet_{\mu\alpha}\omet^{\alpha}_{\nu}]
\end{equation}
and 
\begin{displaymath}
\oric = \oimet^{\mu\nu}r_{\mu\nu} =\frac{3m^2}{2A}(\tau -4A). 
\end{displaymath}
The Lagrangian (\ref{n68}) reads now 
\begin{equation}\label{n76}
L_M = \frac{1}{2}K(Q) +\frac{m^2}{8A}(-\tau^2 + \omet^{\mu\nu}
\omet_{\mu\nu} +6A\tau -12 A^2). 
\end{equation}
Its potential part generates the potential piece of the equations of 
motion (\ref{n71}), 
\begin{equation}\label{n77}
P^{\mu\nu}=\frac{m^2}{8A}\left[
\left(\tau^2-\omet^{\alpha\beta}\omet_{\alpha\beta}-12A^2\right)\oimet^{\mu\nu}
+4\omet^{\mu\nu}+4\left(3A-\tau\right)g^{\mu\nu}\right]
\end{equation} 
with the trace $P^{\mu\nu}\omet_{\mu\nu} = \frac{3}{2}m^2(\tau -4A) =
A\oimet^{\mu\nu}r_{\mu\nu}$. \\

Inserting $r_{\mu\nu}$ from eq.~(\ref{n75}) into the constraints (\ref{n70}) and 
employing an identity valid for any two nonsingular symmetric 
tensors $\omet_{\mu\nu}$ and $g_{\mu\nu}$, 
\begin{equation}\label{n78}
A_{,\mu} \equiv \frac{1}{2}A\oimet^{\alpha\beta}\omet_{\alpha\beta ;\mu}, 
\end{equation} 
one arrives at $J_{\mu}=0$ for $J_{\mu}$ appearing in eq.~(\ref{n73}). The 
constraints $J_{\mu}=0$, which are equivalent to (\ref{n70}), hold 
independently of the equations of motion. Assuming that the equations 
$E^{\mu\nu}=0$ hold one gets from (\ref{n73}) that 
$E^{\mu\nu}\omet_{\mu\nu} = 0 = -P^{\mu\nu}\omet_{\mu\nu}$ and this 
implies that $\tau = 4A$ or $\oric(\omet) =0$. Vanishing of 
$\oric$ is evident from Einstein field equations (\ref{n39}) in HJF. 
Next inserting $A =\tau/4$ in (\ref{n78}) and simplifying the resulting 
equation by means of $J_{\mu}=0$ one gets four differential constraints 
\begin{equation}\label{n79}
\tau_{,\mu} -2\omet_{\mu\alpha}{}^{;\alpha} =0. 
\end{equation} 
Together with the algebraic constraint $\tau=4A$ they form five constraints 
imposed on $\omet_{\mu\nu}$ ensuring that the field has exactly five 
degrees of freedom and carries spin two. As in HJF, the 
gravitational doublet consists of spin--two fields, a massless (the 
metric) and a massive one \cite{ABJT, AEL, HOW1, Tom}, clearly its mass is the same
in both frames.
\\

The algebraic constraint reduces (\ref{n75}) and (\ref{n77}) to 
\begin{equation}\label{n80}
P^{\mu\nu} =\frac{m^2}{2\tau}[(\frac{1}{4}\tau^2 -\omet^{\alpha\beta}
\omet_{\alpha\beta})\oimet^{\mu\nu} +4\omet^{\mu\nu} -
\tau g^{\mu\nu}]
\end{equation}
and
\begin{equation}\label{n81}
r_{\mu\nu} =\frac{m^2}{2\tau}(\tau\omet_{\mu\nu} -4\omet_{\mu\alpha}
\omet_{\nu}^{\alpha}).
\end{equation}
Now one can find a simple relationship between the massive fields in 
both the frames. $\pi^{\mu\nu}$ as a function of the EF variables is 
(from (\ref{n74}) and (\ref{n39}))
\begin{equation}\label{n82}
\pi^{\mu\nu}(g_{\alpha\beta}, \omet_{\alpha\beta}) =
\vert \det(\omet^{\alpha}_{\beta})\vert^{-1/2}g^{\mu\nu} -
\oimet^{\mu\nu},
\end{equation}
we stress that $\oimet^{\mu\nu}$ is the inverse matrix to 
$\omet_{\mu\nu}$ and all indices (including these at $\omet_{\mu\nu}$) 
are shifted with the aid of $g_{\mu\nu}$. \\

By definition $\omet_{\mu\nu}$ is a nonsingular matrix and as such it is 
not the most suitable description of a classical field which should 
vanish in a ground state. A field redefinition is required. To this aim 
we first determine a ground state solution. In flat spacetime a ground 
state solution (vacuum) is Lorentz invariant, in a curvd one it should 
be covariantly constant, $\omet_{\alpha\beta ;\mu} =0$. This implies 
\begin{equation}\label{n83}
Q^{\alpha}_{\mu\nu} =0, \qquad \tau =\textrm{const} \qquad 
\textrm{and} \qquad \omet_{\alpha\beta}\omet^{\alpha\beta} =
\textrm{const}. 
\end{equation}
The equations $E^{\mu\nu} =0$ reduce for this solution to 
$P^{\mu\nu} =0$ and these read 
\begin{equation}\label{n84}
4\omet_{\mu\alpha}\omet_{\nu}^{\alpha} = \tau\omet_{\mu\nu} +
(\omet_{\alpha\beta}\omet^{\alpha\beta} -\frac{1}{4}\tau^2)
g_{\mu\nu}.
\end{equation} 
By inserting (\ref{n84}) into (\ref{n81}) one finds 
\begin{equation}\label{n85}
r_{\mu\nu} = \frac{m^2}{8\tau}(\tau^2 -4\omet_{\alpha\beta}\omet^{\alpha\beta})
g_{\mu\nu} \equiv Cg_{\mu\nu}
\end{equation}
and the energy--momentum tensor is reduced to its potential part 
equal to $T_{\mu\nu} = -Cg_{\mu\nu}$. The ground state spacetime 
satisfies $G_{\mu\nu} = -Cg_{\mu\nu}$ and should be maximally 
symmetric, i.e. Minkowski, dS or AdS depending on the sign of $C$. 
Usually for a classical field its stress $T_{\mu\nu}$ vanishes in the 
ground state; for $\omet_{\mu\nu}$ this amounts to 
$\tau^2 = 4\omet_{\alpha\beta}\omet^{\alpha\beta}$ and $r_{\mu\nu}=0$. 
Multiplying (\ref{n81}) by $\oimet^{\nu\sigma}$ one gets 
$\omet_{\mu\nu} =\frac{1}{4}\tau g_{\mu\nu}$ and $g_{\mu\nu} =
\eta_{\mu\nu}$. Then $A=(\tau/4)^2$ and the constraint $\tau=4A$ yields 
$\tau=4$ as $\tau$ and $A$ cannot vanish. Finally the ground state 
solution is $g_{\mu\nu} =\eta_{\mu\nu} =\omet_{\mu\nu}$. \\
Next assume that $C \ne 0$ and the spacetime is dS or AdS. Any 
covariantly constant $\omet_{\mu\nu}$ can be written as 
$\omet_{\mu\nu}= \frac{1}{4}\tau g_{\mu\nu} +\phi_{\mu\nu}$ with 
$\tau=\textrm{const}$, $g^{\mu\nu}\phi_{\mu\nu}=0$ and 
$\phi_{\mu\nu ;\alpha} =0$. Inserting the Riemann tensor for these two 
spacetimes into the identity 
$R^{\sigma}{}_{\alpha\mu\nu}\phi_{\sigma\beta} +
R^{\sigma}{}_{\beta\mu\nu}\phi_{\sigma\alpha} =0$ which follows from 
Ricci identity for a covariantly constant $\phi_{\mu\nu}$, one 
finds $\phi_{\mu\nu} =0$ --- dS and AdS do not admit a traceless 
covariantly constant $\phi_{\mu\nu}$. Thus we have shown that the theory 
in EF has a unique ground state solution $g_{\mu\nu} =\eta_{\mu\nu} =
\omet_{\mu\nu}$. This result was previously found in \cite{HOW1} under the simplifying
assumption that the spin--two field is proportional to the metric $g_{\mu\nu}$; the
proof here is generic. Of course there are many Einstein spaces in which 
$\omet_{\mu\nu ;\alpha} =0$, e.g. setting $\omet_{\mu\nu} =g_{\mu\nu}$ 
one gets $R_{\mu\nu}(g) =0$ and conversely, for 
$\omet_{\mu\nu ;\alpha} =0$ and $R_{\mu\nu} =0$ the only solution is 
$\omet_{\mu\nu} =g_{\mu\nu}$. However these spacetimes are not maximally 
symmetric and cannot be regarded as ground state solutions.

Hindawi et al.~\cite{HOW1} have shown stability of the ground state solution against
restricted scalar perturbations of the field $\omet_{\mu\nu}$. However to prove
stability of the solution one should show either that arbitrary tensor excitations of
the field do not grow in time, or equivalently that these generic fluctuations have
positive energy density. The second way is impractical taking into account the
complexity of the stress tensor (\ref{n69}): we will return to the problem of energy
density for small excitations in a forthcoming paper. The ground state solution
is mapped by the Legendre transformation (\ref{n56}) onto the
$\omet_{\mu\nu}=\eta_{\mu\nu}$ and $\pi^{\mu\nu}=0$ solution in HJF (sect.~3). The
plane--wave perturbations of the solution in HJF can be transformed with the aid of
the map (\ref{n82}) onto similar perturbations in EF showing the linear stability of
the vacuum in the latter frame.

The massive spin--2 field should be described as an excitation above 
the ground state. In order to have a covariant description one makes 
a redefinition $\omet_{\mu\nu} \equiv g_{\mu\nu} +\Phi_{\mu\nu}$ 
and assumes that the gravitational doublet is EF=$\{g_{\mu\nu}, 
\Phi_{\mu\nu}\}$\footnote{Tomboulis \cite{Tom} employs a different decomposition of
$\omet_{\mu\nu}$ into $g_{\mu\nu}$ and the massive spin--two field.}. In terms of 
$\Phi_{\mu\nu}$ one has 
\begin{displaymath}
Q^{\alpha}_{\mu\nu} =\frac{1}{2}\oimet^{\alpha\beta}
(\Phi_{\beta\mu ;\nu} +\Phi_{\beta\nu ;\mu} -
\Phi_{\mu\nu ;\beta}), \qquad \Phi \equiv g^{\mu\nu}\Phi_{\mu\nu}, 
\end{displaymath}
furthermore $\tau=\Phi +4$ and the constraints read 
\begin{equation}\label{n86}
\Phi +4 = 4\vert\det (\delta^{\alpha}_{\beta} +
\Phi^{\alpha}_{\beta})\vert^{1/2}, \qquad 
\Phi_{\mu\alpha}{}^{;\alpha} -\frac{1}{2}\Phi_{,\mu} =0.  
\end{equation}
The field equations explicitly expressed in terms of $\Phi_{\mu\nu}$ 
are given in Appendix. 

\section{Internal structure of the theory in EF}

Consistency of the field equations in EF follows from the consistency 
in HJF and dynamical equivalence of the two frames. Alternatively, 
one can prove it directly in EF using, as previously, the 
coordinate invariance. Now the proof is slightly different. Under an 
infinitesimal transformation $x'^{\mu} =x^{\mu} +\varepsilon^{\mu}$ 
one finds $\delta\Phi_{\mu\nu} =2\eta_{(\mu ;\nu)} +
2\varepsilon_{\alpha}\Omega_{\mu\nu}{}^{\alpha}$, where 
$\eta_{\mu} \equiv -\Phi_{\mu\alpha}\varepsilon^{\alpha}$ and 
$\Omega_{\mu\nu}{}^{\alpha} \equiv \frac{1}{2}
(\Phi^{\alpha}{}_{\mu ;\nu} + \Phi^{\alpha}{}_{\nu ;\mu} -
\Phi_{\mu\nu}{}^{;\alpha})$. Using the definition (\ref{n71}) in 
$\delta \int \ud^4 x\sqrt{-g}L_M =0$ one arrives at an analogous 
strong Noether conservation law, 
\begin{equation}\label{n87}
T^{\mu\nu}{}_{;\nu} -E^{\alpha\beta}{}_{;\beta}\Phi^{\mu}_{\alpha} -
\Omega_{\alpha\beta}{}^{\mu}E^{\alpha\beta} =0.
\end{equation}
The energy--momentum tensor has a purely geometric origin, i.e. 
setting $\oric_{\mu\nu} =r_{\mu\nu}$ in (\ref{n69}) one gets 
$T_{\mu\nu} =G_{\mu\nu}$, thus its divergence always vanishes, 
$T^{\mu\nu}{}_{;\nu} =0$. Consistency then requires that 
$E^{\alpha\beta}{}_{;\beta} =0$ identically if $E^{\alpha\beta} =0$ 
and the five constraints hold in the spacetime. However eqs.~(\ref{n71}) 
are too complicated to allow for a direct check of the identity. 
This will be done in a subsequent paper in a perturbative analysis 
about the ground state solution up to the second order. \\

Also the problem of whether the Lagrange equations for $\Phi_{\mu\nu}$ 
are all hyperbolic propagation ones is harder in this frame. Here 
one should distinguish between eqs.~(\ref{n71}) and (\ref{ap2}) since the latter 
arise from the former upon using the constraints. After long 
computations one finds that all eqs.~(\ref{n71}) contain second time 
derivatives $\Phi_{\mu\nu,00}$; this is due to the nonlinearities. 
It is difficult to establish whether these equations are hyperbolic. 
We will return to the problem in a subsequent paper where it will 
be shown that up to the second order in an perturbation expansion 
these form a nondegenerate system of hyperbolic propagation 
equations (of Klein--Gordon type). \\

At present we investigate the structure of the linearized theory. Let 
a weak field $\Phi_{\mu\nu}$ be the source of some metric $g_{\mu\nu}$. 
One should not a priori assume that $g_{\mu\nu}$ is a fixed 
spacetime background for a small excitation $\Phi_{\mu\nu}$ since 
$T_{\mu\nu}(g, \Phi)$ may contain linear terms and the metric will 
then be affected by the excitations. The linearized form of eqs.~(\ref{n71}) 
reads 
\begin{eqnarray}\label{n88}
2E^{(1)}_{\mu\nu} & = &
g_{\mu\nu}\left(\Phi^{\alpha\beta}{}_{;\alpha\beta}-\Box\Phi\right)
+\Box\Phi_{\mu\nu}-R^{\alpha}{}_{(\mu}\Phi_{\nu)\alpha}
-\Phi^{\alpha}{}_{\mu;\alpha\nu}-\Phi^{\alpha}{}_{\nu;\alpha\mu}\nonumber\\
&& +2R_{\mu\alpha\nu\beta}\Phi^{\alpha\beta}+\Phi_{;\mu\nu}
-m^2\left(\Phi_{\mu\nu}-\Phi g_{\mu\nu}\right)=0,
\end{eqnarray} 
here we have used $A \approx 1 +\frac{1}{2}\Phi$. The algebraic 
constraint gives in the linear approximation $\Phi =0$ and the 
differential ones reduce to $\Phi_{\mu\nu}{}^{;\nu} =0$. Applying 
these constraints one gets the linearized version of eqs.~(\ref{ap2}) which 
one denotes by $E^{\textrm{L}}_{\mu\nu} =0$. It turns out that the 
linearized energy--momentum tensor (\ref{ap1}) is 
$T^{\textrm{L}}_{\mu\nu} =E^{\textrm{L}}_{\mu\nu}$, thus for the 
linearized equations of motion (\ref{ap1}--\ref{ap2}) one gets 
$T^{\textrm{L}}_{\mu\nu} =0$ and $R_{\mu\nu}(g) =0$ (thus 
$\Phi_{\mu\nu}$ is actually decoupled from the metric which becomes a 
background) and finally $E^{\textrm{L}}_{\mu\nu}$ is simplified to 
\begin{equation}\label{n89}
2E^{\textrm{L}}_{\mu\nu} =\Box\Phi_{\mu\nu} -m^2\Phi_{\mu\nu} +
2R_{\mu\alpha\nu\beta}\Phi^{\alpha\beta} =0;
\end{equation}
clearly $E^{(1)}_{\mu\nu}$ coincides with $E^{\textrm{L}}_{\mu\nu}$ 
upon employing the constraints. Eq.~(\ref{n89}) is identical to the linearized 
form of eq.~(\ref{n40}) for $\pi^{\mu\nu}$ in HJF upon replacing $g_{\mu\nu}$ 
by $\omet_{\mu\nu}$. This follows from eq.~(\ref{n82}) where one puts 
$A \approx 1$ and $\oimet^{\mu\nu} \approx g^{\mu\nu} -\Phi^{\mu\nu}$. 
Then $\Phi^{\mu\nu} \approx \pi^{\mu\nu}$   and in this approximation 
one may replace $\nabla_{\mu}$ by $\onab_{\mu}$. \\

Now one should show that the constraints $\Phi = 0 = 
\Phi_{\mu\nu}{}^{;\nu}$ (which for the moment will be referred to as 
secondary constraints) are preserved in time by the linear equations 
(\ref{n89}) and $R_{\mu\nu}(g) =0$. To this end one determines which 
equations of the system (\ref{n88}) are not propagation ones. Using the 
Gauss normal coordinates in which $g_{00}=-1$ and $g_{0i}=0$, 
$i =1, 2,3$, one finds that the four equations $E^{(1)}_{0\mu} =0$ 
(and equivalently $E^{(1)0\mu} =0$ and $E^{(1)0}_{\mu} =0$) do not 
contain the time derivatives $\Phi_{\mu\nu,00}$; these will be 
referred to as primary constraints. Hence eqs.~(\ref{n88}) form a degenerate 
system consisting of 6 propagation equations $E^{(1)}_{ik} =0$ and 
the four primary constraints. This allows one to prove a proposition: \\

{\sl If the propagation equations (\ref{n89}) and $R_{\mu\nu}(g) =0$ hold 
throughout the spacetime and the following constraints restrict 
initial data on a given Cauchy surface: 
\begin{itemize}
\item the four primary constraints $E^{(1)}_{0\mu} =0$, 
\item the five secondary constraints $\Phi = 0 = 
\Phi_{\mu\nu}{}^{;\nu}$ and 
\item additionally $\Phi_{,0} =0$, 
\end{itemize} 
then all the constraints, $\Phi = 0 = \Phi_{\mu\nu}{}^{;\nu}$ and 
$E^{(1)}_{0\mu} =0$, are preserved in time and the eqs.(\ref{n89}) are 
equivalent to (\ref{n88}).} \\

\noindent {\bf Proof:} 
Let the Cauchy surface have a local equation $x^0 =0$ in Gauss normal 
coordinates. The additional condition  $\Phi_{,0} =0$ at $x^0 =0$ is 
essential since there are less primary constraints than the secondary 
ones. \\ 
a) The constraint $\Phi =0$. A propagation equation for $\Phi$ is 
provided by taking the trace of eq.~(\ref{n89}), 
$2g^{\mu\nu}E^{\textrm{L}}_{\mu\nu} =\Box\Phi -m^2 \Phi =0$ since 
$R_{\alpha\beta} =0$. The unique solution satisfying the initial 
conditions is $\Phi =0$. \\
b) A propagation equation for the vector 
$S_{\mu} \equiv \Phi_{\mu\nu}{}^{;\nu}$ arises by taking divergence of 
eq.~(\ref{n89}). In fact, in empty ($R_{\mu\nu} =0$) spacetime the 
contraction of the full Bianchi identity yields 
$R_{\alpha\beta\mu\nu}{}^{;\nu} =0$, then 
$\nabla^{\nu}\Box\Phi_{\mu\nu} =\Box S_{\mu} +
2R_{\sigma\mu\alpha\nu}\Phi^{\sigma\nu ;\alpha}$ and 
$2E^{\textrm{L}}_{\mu\nu}{}^{;\nu} =0$ is reduced to the vector 
Klein--Gordon equation, $ \Box S_{\mu} -m^2S_{\mu} =0$. Next 
applying $\Phi =0$ one reexpresses the primary constraints in (\ref{n88}) 
in terms of $S_{\mu}$, $2E^{\textrm{(1)}}_{0\mu}=g_{0\mu}S_{\alpha}{}^{;\alpha}-
S_{0;\mu}-S_{\mu;0}=0$ at $x_0=0$. In Gauss normal coordinates these read $S_{\mu,0}=0$.
The initial data $S_{\mu}=S_{\mu,0}=0$ at $x_0=0$ imply then that $S_{\mu}\equiv 0$ for
all times. This in turn implies vanishing of $E^{\textrm{(1)}}_{0\mu}$ in spacetime,
furthermore eqs.~(\ref{n89}) and (\ref{n88}) become equivalent everywhere.\\

As a byproduct it has been shown that the trace and divergence of eqs.~(\ref{n89})
vanish identically giving rise to no further constraints besides the five secondary
ones. This shows that the linear theory is fully consistent. It should be stressed
that this theory is not identical with that for the linear massive test spin--2
field in empty spacetime, \cite{AD1,AD2}, though both fields are subject to the same
equations of motion, (\ref{n89}) and $R_{\mu\nu}=0$, and to the same constraints. 
In fact, for the linear Fierz-Pauli
spin--2 field $\Psi_{\mu\nu}$, it is known that all possible Lagrangians are
equivalent (in flat spacetime) to the Wentzel Lagrangian
\begin{eqnarray}\label{n90}
L_W(\Psi,m)&\equiv&\frac{1}{4}\left(-\Psi^{\mu\nu,\alpha}\Psi_{\mu\nu,\alpha}
+2\Psi^{\mu\nu,\alpha}\Psi_{\alpha\mu,\nu}
-2\Psi^{\mu\nu}{}_{,\nu}\Psi_{,\mu}+\Psi^{,\mu}\Psi_{,\mu}\right)\nonumber\\&&
-\frac{m^2}{4}\left(\Psi^{\mu\nu}\Psi_{\mu\nu}-\Psi^2\right),
\end{eqnarray}
where $\Psi\equiv\eta^{\mu\nu}\Psi_{\mu\nu}$ \cite{Wen,AD1,AD2,HOR,MS3}.
In contrast, the linearized theory for the field $\Phi_{\mu\nu}$ above is not
self--contained as a theory for a free spin--two field in a fixed background (in
particular, it has not its own Lagrangian) and arises only as a limit case of the
nonlinear theory based on the Helmholtz Lagrangian (\ref{n66}) and (\ref{n76}). 

The difference is most easily seen while dealing with energy density. The
Wentzel Lagrangian generates a quadratic stress tensor $T^W_{\mu\nu}$ for the linear
field \cite{AD2,MS3}, while in the linearized theory for $\Phi_{\mu\nu}$ one finds that
the lowest order terms of the expansion of the stress tensor are linear and vanish
identically for solutions of the linearized field equations, while the quadratic part
in the expansion of $T_{\mu\nu}$ (which is not presented here) does \emph{not} coincide
with $T^W_{\mu\nu}$.

This fact seems surprising, but one has to keep in mind that $T^W_{\mu\nu}$
is derived from $L_W$ by replacing the fixed metric
$\eta_{\mu\nu}$ by a generic metric $g_{\mu\nu}$ and the ordinary derivatives with
covariant ones, then taking the variational derivative with respect to $g_{\mu\nu}$. In
this derivation one regards the variation of the spin--two field to be zero,
$\delta_g\Phi_{\mu\nu}\equiv 0$. On the contrary, when computing the stress tensor for
the exact Lagrangian $L_M$ (\ref{n68}), one regards the field $\omet_{\mu\nu}$ as
fundamental, $\delta_g\omet_{\mu\nu}\equiv 0$, and since $\omet_{\mu\nu} \equiv
g_{\mu\nu} +\Phi_{\mu\nu}$ then $\delta_g\Phi_{\mu\nu}= -\delta
g_{\mu\nu}$. The fact that the spin--two field excitations are defined to vanish when
$\omet_{\mu\nu}$ is equal to the actual spacetime metric $g_{\mu\nu}$ (and not to some
fixed background metric, which would be physically objectionable and would also make the
equations rather cumbersome) implies that the true stress tensor will contain both
linear and quadratic terms not related to the Wentzel Lagrangian. Furthermore, even
assuming $\delta_g\Phi_{\mu\nu}= -\delta g_{\mu\nu}$ while taking the variation of $L_W$
would not be enough to recover the quadratic terms of the true stress tensor, because
the higher order terms in the expansion of $L_M$ do contribute to quadratic terms in the
stress tensor. The case of the linearized theory for
$\Phi_{\mu\nu}$ clearly shows that in general the equations of motion alone are
insufficient for determining an energy--momentum tensor for a given field \cite{MS2}.

Finally we find the expression for the Helmholtz Lagrangian (\ref{n66}) and
(\ref{n76}) in the lowest order approximation around the ground state solution
$g_{\mu\nu}=\eta_{\mu\nu}$ and $\Phi_{\mu\nu}=0$
(where \emph{both} metric and massive field perturbations are now taken into account).
The metric and massive field excitations are $g_{\mu\nu}=\eta_{\mu\nu}+\epsilon
h_{\mu\nu}$ and $\Phi_{\mu\nu}=\epsilon\varphi_{\mu\nu}+\epsilon^2
\xi_{\mu\nu}$. Expanding the
gravitational part one finds $R(g)\sqrt{-g}\approx \epsilon^2 L_W(h,0)$ (the Wentzel
Lagrangian is in fact known to coincide with the lowest--order expansion of the
Einstein--Hilbert Lagrangian),  and similarly for
$2L_M$ in eq.~(\ref{n76}), so that up to a full divergence (some auxiliary expressions
are given in Appendix)
\begin{equation}\label{n91}
L_H(\eta_{\mu\nu}+\epsilon
h_{\mu\nu},\epsilon\varphi_{\mu\nu}+\epsilon^2
\xi_{\mu\nu})\sqrt{-\det(\eta_{\mu\nu}+\epsilon h_{\mu\nu})}\approx
\epsilon^2\left[L_W(h,0)-L_W(\varphi,m)\right].
\end{equation}
Up to the second order the fields are decoupled (there are no interaction terms); also
the field $\xi_{\mu\nu}$ is absent. In this way we have rederived the well known fact
that in the linear approximation the massive spin--two field $\varphi_{\mu\nu}$ arising
from a nonlinear gravity theory is a ghost field (a ``poltergeist'')
\cite{St1,St2,ABJT,Tom,HOW1,HOW2} (a full family of unphysical ghost fields
appears in the linearized theory according to a different approach in \cite{BJS}).
This outcome is inescapable since Wald \cite{Wald2} gave a generic argument that in any
generally covariant theory of a number of consistently interacting spin--two fields at
least one field is necessarily ghostlike. Thus consistency of gravitational interactions
implies that the massive spin--two field produces states of negative norm in the state
vector space of quantum theory.

\section{Exact solutions in Einstein frame}

The field equations (\ref{ap1}--\ref{ap2}) are fairly involved, nevertheless one may seek for
nontrivial solutions. It is well known (cf.~e.g.~\cite{St1}) that the corresponding
fourth--order field equations for the metric $\tilde{g}_{\mu\nu}$ in JF admit solutions
$\oric_{\mu\nu}(\tilde{g})=0$ which are trivial in the sense that they are vacuum
solutions already in Einstein's general relativity. We therefore regard as nontrivial
those solutions to eqs.~(\ref{ap1}--\ref{ap2}) which after transforming back to JF yield
nonvanishing Ricci tensor.

A general method for seeking solutions is to investigate known classes of geometrically
distinguished metrics depending on some arbitrary functions and check if they can satisfy
Einstein field equations (\ref{ap1}) for a suitably chosen field $\Phi_{\mu\nu}$. If
solutions are not precluded (see below) one may attempt to solve the entire system
(\ref{ap1}--\ref{ap2}). Clearly some simplifying assumptions regarding $\Phi_{\mu\nu}$ are
indispensable. It turns out that the constraints (\ref{n86}) are very stringent and one
should solve them for a given class of $\Phi_{\mu\nu}$ and $g_{\mu\nu}$ before dealing
with the field equations.

A generic simplifying assumption is to represent the tensor $\Phi_{\mu\nu}$ in terms of
few scalar or vector functions. The simplest ansatz, that $\Phi_{\mu\nu}$ is completely
described by its trace, $\Phi_{\mu\nu}=\frac{1}{4}\Phi g_{\mu\nu}$, is excluded by the
constraints. In fact, the algebraic constraint (\ref{n86}) is then easily solved yielding
$\Phi=-4$ or $\Phi=0$. The first solution is excluded because it would correspond to 
vanishing of the original metric, $\tilde{g}_{\mu\nu}=0$, while the second is trivial (in
the above sense) since for $\Phi_{\mu\nu}=0$ one
has $R_{\mu\nu}(g)=\oric_{\mu\nu}(\tilde{g})=0$.

A static spherically symmetric solution is not known. Then we consider the simpler case
of a metric representing a plane--fronted gravitational wave with parallel rays (a
\emph{pp wave}) \cite{EK, Kr}. Such waves are characterized by a null covariantly
constant wave vector (ray) $k_{\mu}$, $k^{\mu}k_{\mu}=0$ and $k_{\mu;\nu}=0$. First
we assume that the spin--two field is of the form
$\Phi_{\mu\nu}=2k_{(\mu}W_{,\nu)}$, where $W$ is some scalar function, and it is null in
the sense that $\Phi_{\mu\nu}k^{\nu}=0$. The latter condition holds if
$k^{\mu}W_{,\mu}=0$, what implies that the gradient $W_{,\mu}$ is either null and
proportional to $k_{\mu}$, or it is spacelike and orthogonal to $k_{\mu}$. If we
consider the spacelike case, $W^{,\mu}W_{,\mu}>0$, we immediately get the trace $\Phi=0$
and then the algebraic constraint holds identically for any scalar $W$. The four
differential constraints (\ref{n86}) reduce to the equation $\Box W=0$. Then one finds
that the expression of the stress tensor becomes
\begin{equation}\label{n93}
T_{\mu\nu}= -m^2\left(k_{(\mu}W_{,\nu)}+\frac{1}{2}W^{,\alpha}W_{,\alpha}k_{\mu}k_{\nu}
\right),
\end{equation}
which is traceless. However, it is known \cite{Kr} that Ricci tensor for a pp wave is
proportional to $k_{\mu}k_{\nu}$. This condition is met by $T_{\mu\nu}$ as in (\ref{n93})
only if $W^{,\mu}W_{,\mu}=0$, contrary to our assumption. 

The other possibility is $\Phi_{\mu\nu}=w\,k_{\mu}k_{\nu}$ (which
includes the special case of the previous ansatz where $W_{,\mu}=w k_{\mu}$). Once again 
$\Phi=0$ and the algebraic constraint becomes an identity. The field $\Phi_{\mu\nu}$ is
null by definition and the condition $k^{\mu}w_{,\mu}=0$ arises now from the
differential cosntraints. The inverse matrix is
$\oimet^{\mu\nu}=g^{\mu\nu}-\Phi^{\mu\nu}$, while 
\begin{equation}\label{n94}
Q^{\alpha}_{\mu\nu}=k^{\alpha}k_{(\mu}w_{,\nu)}-\frac{1}{2}w^{,\alpha}k_{\mu}k_{\nu}
\end{equation}
implies
\begin{equation}\label{n95}
Q^{\alpha}_{\mu\nu;\alpha}=-\frac{1}{2}k_{\mu}k_{\nu}\Box w.
\end{equation}
The energy--momentum tensor, 
\begin{equation}\label{n96}
T_{\mu\nu}= \frac{1}{2}k_{\mu}k_{\nu}\left(\Box w-m^2w\right),
\end{equation}
is admissible by the pp--wave metric; one then evaluates the Lagrange equations of
motion  (\ref{ap2}): these reduce to one scalar equation
\begin{equation}\label{n97}
\left(\Box -m^2\right)w=0,
\end{equation}
hence $T_{\mu\nu}=0$ and $R_{\mu\nu}=0$. We have arrived at a rather surprising result:
although the spacetime is not empty since $\Phi_{\mu\nu}\ne 0$, the metric satisfies the
vacuum field equations and the spin--two field necessarily behaves as a test matter and
carries no energy. To elucidate it one takes into account that a pp--wave metric
represents its own linear approximation around flat spacetime \cite{EK,GR,G}: the
inverse matrix $g^{\mu\nu}$ equals its linearized version and the exact Einstein
equations become linear and coincide with the linearized Einstein tensor. In this case
linearity of the equations of motion and the assumption that the presence of the massive
field preserves the pp--wave form of the metric imply that the interaction of the metric
with its source is quite restricted. In fact, $\Phi_{\mu\nu}=w\,k_{\mu}k_{\nu}$ means
that the massive field propagates in the same direction as the pp wave or that their
momenta are parallel. In this situation the equations of motion and $T_{\mu\nu}$ are
reduced to their linear parts. As we have seen in sect.~7, in the linearized theory
$T^{L}_{\mu\nu}=0$ by virtue of $E^{L}_{\mu\nu}=0$ for any metric, not necessarily being
a pp wave.

To complete the solution one solves (\ref{n97}) for $w$. Writing the pp--wave metric as
in \cite{Kr}
\begin{equation}\label{n98}
ds^2= -2H(u,x,y)du^2-2du\,dv + dx^2 + dy^2,
\end{equation}
where the hypersurfaces $u=$const are null and are wave--fronts and
$k_{\mu}=-\partial_{\mu}u$, one finds $\frac{\partial w}{\partial v}=0$ and the
Klein--Gordon eq.~(\ref{n97}) is reduced to Yukawa equation on Euclidean plane,
\begin{equation}\label{n99}
\frac{\partial^2 w}{\partial x^2}+\frac{\partial^2 w}{\partial y^2}-m^2 w =0.
\end{equation}
Solutions to the latter one are well known, and employing polar coordinates $r$ and
$\theta$ are expressed in terms of Bessel functions
\begin{eqnarray}\label{n100}
w(u,r,\theta) &=& \left[A_0(u)J_0(imr)+iB_0(u)H^{(1)}_0(imr)\right]
\left[C_0(u)\theta+D_0(u)
\right]\\&&+
\sum_{n=1}^{\infty}\left[A_n(u)I_n(imr)+B_n(u)K_n(imr)\right]
\left[C_n(u)\sin(n\theta)+D_n(u)\cos(n\theta)\right],\nonumber
\end{eqnarray}
here $I_n$ and $K_n$ are the modified Bessel functions, $J_0$ and $H^{(1)}_0$ are the
usual Bessel and Hankel functions, while $A_n$, $B_n$, $C_n$ and $D_n$
($n=0,\ldots\infty$) are arbitrary functions of the retarded time $u$.

It should be emphasized that the solution (\ref{n96})--(\ref{n100}) is nontrivial in
spite of $R_{\mu\nu}=0$. To show it one employs eq.~(\ref{n74}); first its trace w.r.~to 
$\omet_{\mu\nu}$ yields $\oric=0$, then the transformation is reversed to give
\begin{equation}\label{n101}
\oric^{\mu\nu}=-\frac{m^2}{2}w\,k^{\mu}k^{\nu}.
\end{equation}
The ray vector is defined by $k_{\mu}=-\partial_{\mu}u$ and then
$k^{\mu}=g^{\mu\nu}k_{\nu}=\oimet^{\mu\nu}k_{\nu}$, thus it is null in both metrics, 
$g^{\mu\nu}k_{\mu}k_{\nu}=\oimet^{\mu\nu}k_{\mu}k_{\nu}=0$. This stems from the fact
that $\omet_{\mu\nu}$ is a pp--wave metric too of the form (\ref{n98}), with the
determining function $\tilde{H}(u,x,y)=H-\frac{1}{2}w$. From (\ref{n98}) one gets 
$R_{\mu\nu}=k_{\mu}k_{\nu}\Box H$, where $\Box H$ reduces to the 2--dimensional
Laplacian as in (\ref{n99}) and $\Box H=0$ for $R_{\mu\nu}=0$. In Jordan frame one has
$\obox\tilde{H}=\Box\tilde{H}$ and (\ref{n101}) follows. One checks directly that
eq.~(\ref{n101}) represents a solution of the fourth-order field equations. The ray
vector is covariantly constant w.r.~to $\omet_{\mu\nu}$ too, $\onab_{\mu}k_{\nu}=0$,
since this property is independent of the metric functions $H$ and $\tilde{H}$. 

Clearly in HJF Ricci tensor is also given by eq.~(\ref{n101}). Then, according to
(\ref{n35}), $\pi_{\mu\nu}=w\,k_{\mu}k_{\nu}$ and its energy--momentum tensor is equal,
by virtue of (\ref{n36}), to Ricci tensor. At first sight it seems better than in EF
because here a non--vanishing field carries a non--vanishing energy. However a stress
tensor of the form (\ref{n101}) describes a flux of energy and momentum moving at light
velocity, as e.g. for a plane monochromatic electromagnetic wave. This is not the case of
a massive field (particle) of any spin. This stress tensor in HJF is misleading and in
this sense the vanishing of $T_{\mu\nu}$ in EF is closer to reality. However it should
be emphasized that this bizarre situation is due to the peculiar choice
$\Phi_{\mu\nu}=w\,k_{\mu}k_{\nu}$. One should compare the stress tensors in both frames
for more physical solutions.

\section{Conclusions}

A model for a massive spin--two field consistently interacting with Einstein's gravity
is provided by a nonlinear gravity theory, in particular by its simplest version with a
Lagrangian quadratic in the Ricci tensor. The consistency is achieved at the cost of
introducing the nonlinearity. The model can be formulated as a second order Lagrangian
field theory in many mathematically equivalent versions (``frames''). Among them two are
distinguished: HJF and EF. In both frames the respective spacetime metric satisfies
Einstein's field equations with the spin--two field acting as a matter source: this is in
fact a result of the Legendre transformation, not of the choice of a particular
frame. 

Though the equations of motion in both frames are similar, the Lagrangian
structure is different. While in EF the model has the standard form of classical field
theory, in HJF the essential nonminimal coupling to curvature and the absence of a
kinetic Lagrangian for the spin--two field indicate that the field should be viewed as a
nongeometric member of the gravitational doublet (or triplet) rather that as ordinary
matter.

The transformation between the two frames is akin (in a generalized sense) to the known
canonical transformation of Hamiltonian mechanics which interchanges the roles of
particles' positions and momenta, $(q,p)\mapsto(-p,q)$. In a similar way, one cannot
assume that both the old and the new variables have the same physical interpretation as
measurable quantities: in most cases one set of variables is in this sense unphysical.
It is well known that for a quadratic Lagrangian (\ref{n1}) the total gravitational
energy (in JF) is indefinite \cite{Bo,BDS} and in HJF the energy--momentum tensor for
the massive spin--two field is unphysical as being purely linear. Thus HJF is unphysical
in both senses, while at least in the case of restricted NLG theories ($L=f(\oric)$)
Einstein frame is physical since the positive energy theorem holds. Thus energy plays a
crucial role in determining the physical (measurable) variables because energy density is
qualitatively sensitive to the change of the spacetime metric between HJF and EF.
Furthermore Einstein frame is advantageous in that it is unique while HJF is not.

It is not quite clear how harmful for classical nonlinear field theory is the fact that
in the lowest order approximation the massive spin--two field is a ghost. The (exact)
ground state solution is stable against linear perturbations. As regards energy, one
cannot expect that the true expression for the energy--momentum tensor for the field
(that in EF) is positive definite. That this is not the case follows from the existence
of the solution found in sect.~8 for which $T_{\mu\nu}=0$. This energy--momentum
tensor violates the commonly accepted postulate that $T_{\mu\nu}$ vanishes if and only if
the matter field vanishes \cite{HE}. Nevertheless one may still regard this model as a
viable theory for a massive spin--two field. To do justice to the model one should study
it in the second--order approximation. This will be done in a forthcoming paper.

\section*{Acknowledgements}
L.M.S.~is grateful for the hospitality at the Dipartimento di Matematica of
the University of Torino, where the greater part of the work was done. The
work of L.M.S.~is supported in part by KBN grant no. 2P03D01417.

\section*{Appendix}
From the constraints (\ref{n86}) one derives $Q_{\mu\alpha}^{\alpha}
=\Phi_{,\mu}\,(\Phi+4)^{-1}$ and $g^{\mu\nu}Q_{\mu\nu}^{\alpha}=0$.
Then Einstein field equations are
\begin{eqnarray}\label{ap1}
G_{\mu\nu}(g) &=& T_{\mu\nu}(g,\Phi) =-Q_{\mu\nu;\alpha}^{\alpha}
-Q_{\mu\nu}^{\alpha}\Phi_{,\alpha}(\Phi+4)^{-1}
+Q^{\alpha}_{\mu\beta}Q^{\beta}_{\alpha\nu}
-\frac{1}{2}g_{\mu\nu}g^{\alpha\beta}Q^{\sigma}_{\alpha\tau}Q^{\tau}_{\sigma\beta}\nonumber\\
&& 
+\left(\Phi_{;\mu\nu}-\frac{1}{2}g_{\mu\nu}\Box\Phi\right)(\Phi+4)^{-1}
-\left(\Phi_{,\mu}\Phi_{,\nu}
-\frac{1}{2}g_{\mu\nu}\Phi_{,\alpha}\Phi^{,\alpha}\right)(\Phi+4)^{-2}\\
&&
+\frac{m^2}{2}\left[(\Phi-4)\Phi_{\mu\nu}-4\Phi_{\mu\alpha}\Phi_{\nu}{}^{\alpha}
-\frac{1}{2}g_{\mu\nu}\left(\Phi^2-2\Phi-4\Phi_{\alpha\beta}\Phi^{\alpha\beta}\right)
\right](\Phi+4)^{-1},\nonumber
\end{eqnarray}
this energy--momentum tensor is much simpler than that for a linear massive spin--two
field in Minkowski space \cite{AD2}. The Lagrange field equations read
\begin{eqnarray}\label{ap2}
E^{\mu\nu} & = & 
\oimet^{\alpha(\mu}\nabla^{\nu)}\left(\Phi_{,\alpha}\,(\Phi+4)^{-1}\right)
+\oimet^{\alpha\beta}\left[
-\frac{1}{2}g^{\mu\nu}\nabla_{\alpha}\left(\Phi_{,\beta}\,(\Phi+4)^{-1}\right)
+2\nabla_{\alpha}\left(\oimet^{\lambda(\mu}\Phi^{\nu)}{}_{[\lambda;\beta]}\right)\right.\nonumber\\
&&
\left.-2\oimet^{\tau\sigma}\oimet^{\lambda(\mu}\Phi^{\nu)}{}_{[\lambda;\tau]}
\Phi_{\alpha\sigma;\beta}+
\left(
\frac{1}{2}g^{\mu\nu}\oimet^{\lambda\sigma}\Phi_{\alpha\lambda;\sigma}
-\oimet^{\lambda(\mu}\Phi_{\alpha\lambda}{}^{;\nu)}
\right)
\Phi_{,\beta}\,(\Phi+4)^{-1}\right]
\nonumber\\
&&
+\oimet^{\alpha(\mu}\oimet^{\nu)\beta}\left[
-\frac{1}{2}\Box\Phi_{\alpha\beta}+\oimet^{\lambda\rho}\left(
\frac{1}{2}\Phi_{\alpha\lambda}{}^{;\sigma}\Phi_{\beta\rho;\sigma}
+2\Phi^{\sigma}{}_{[\alpha;\lambda]}
\Phi_{\sigma[\beta;\rho]}
\right)
\right]\nonumber\\
&&
-\frac{m^2}{2}\left[\left(\frac{1}{4}
\Phi^2-\Phi_{\alpha\beta}\Phi^{\alpha\beta}\right)
\oimet^{\mu\nu}+4\Phi^{\mu\nu}-\Phi g^{\mu\nu}
\right](\Phi+4)^{-1}=0,
\end{eqnarray}
where $\Box\Phi_{\alpha\beta}\equiv\Phi_{\alpha\beta;\lambda}{}^{;\lambda}$.

In expanding the Helmholtz Lagrangian density $L_H\sqrt{-g}$ up to second order around
the ground state solution the following expressions are useful,
\begin{equation}\label{ap3}
K(Q)\sqrt{-g}\approx-\epsilon^2 L_W(\varphi,0),
\end{equation}
\begin{equation}\label{ap4}
\Phi\approx \epsilon\varphi + \epsilon^2(\xi-h^{\mu\nu}\varphi_{\mu\nu}),\qquad
\Phi^{\mu\nu}\Phi_{\mu\nu}\approx\epsilon^2\varphi^{\mu\nu}\varphi_{\mu\nu},
\end{equation}
where $\varphi\equiv\eta^{\mu\nu}\varphi_{\mu\nu}$ and
$\xi\equiv\eta^{\mu\nu}\xi_{\mu\nu}$,
\begin{equation}\label{ap5}
A\approx 1+\frac{1}{2}\epsilon\varphi+\frac{1}{2}\epsilon^2\left(
\frac{1}{4}\varphi^2-\frac{1}{2}\varphi_{\alpha\beta}\varphi^{\alpha\beta}
-\varphi_{\alpha\beta}h^{\alpha\beta}+\xi
\right),
\end{equation}
\begin{equation}\label{ap6}
\tau\approx 4+\epsilon\varphi+\epsilon^2\left(\xi-\varphi_{\alpha\beta}h^{\alpha\beta}
\right)
\end{equation}
and
\begin{equation}\label{ap7}
\psi^{\alpha\beta}\psi_{\alpha\beta}\approx
4+2\epsilon\varphi+\epsilon^2\left(2\xi-2\varphi_{\alpha\beta}h^{\alpha\beta}+
\varphi_{\alpha\beta}\varphi^{\alpha\beta}\right).
\end{equation}

\addtolength{\baselineskip}{-1pt}

\end{document}